\documentclass{article}

\usepackage{arxiv}

\usepackage[utf8]{inputenc} 
\usepackage[T1]{fontenc}    
\usepackage{hyperref}       
\usepackage{url}            
\usepackage{booktabs}       
\usepackage{amsfonts}       
\usepackage{nicefrac}       
\usepackage{microtype}      
\usepackage{lipsum}
\usepackage{color}
\usepackage{algorithm}
\usepackage{algorithmic}
\usepackage{amsmath,amsfonts,amssymb}
\usepackage{subfigure}
\usepackage{lineno}
\usepackage{graphicx}
\usepackage{bm}
\graphicspath{{Figures/}}
\usepackage{subfigure}

\definecolor{green}{rgb}{0.2, 0.55, 0.02}

\title{Probabilistic neural networks for fluid flow model-order reduction and data recovery}

\author{
  Romit Maulik \\
  Argonne Leadership Computing Facility \\
  Argonne National Laboratory \\
  Lemont, IL 60439 \\
  \texttt{rmaulik@anl.gov} \\
  \And
  Kai Fukami\\
  School of Science for Open and Environmental System \\
  Keio University \\
  Yokohama, 223-8522, Japan \\
  \texttt{kai.fukami@keio.jp}
  \And
  Nesar Ramachandra \\
  High Energy Physics Division \\
  Argonne National Laboratory \\
  Lemont, IL 60439 \\
  \texttt{nramachandra@anl.gov}
  \And
  Koji Fukagata\\
  Mechanical Engineering \\
  Keio University \\
  Yokohama, 223-8522, Japan \\
  \texttt{fukagata@mech.keio.ac.jp}
  \And
  Kunihiko Taira\\
  Mechanical and Aerospace Engineering \\
  University of California, Los Angeles\\
  CA 90095\\
  \texttt{ktaira@seas.ucla.edu}
}

\begin{document}
\maketitle

\begin{abstract}
We consider the use of probabilistic neural networks for fluid flow {surrogate modeling} and data recovery. This framework is constructed by assuming that the target variables are sampled from a Gaussian distribution conditioned on the inputs. Consequently, the overall formulation sets up a procedure to predict the hyperparameters of this distribution which are then used to compute an objective function given training data. We demonstrate that this framework has the ability to provide for prediction confidence intervals based on the assumption of a probabilistic posterior, given an appropriate model architecture and adequate training data. The applicability of the present framework to cases with noisy measurements and limited observations is also assessed. To demonstrate the capabilities of this framework, we consider canonical regression problems of fluid dynamics from the viewpoint of reduced-order modeling and spatial data recovery for four canonical data sets. The examples considered in this study arise from (1) the shallow water equations, (2) a two-dimensional cylinder flow, (3) the wake of NACA0012 airfoil with a Gurney flap, and (4) the NOAA sea surface temperature data set. The present results indicate that the probabilistic neural network not only produces a machine-learning-based fluid flow {surrogate} model but also systematically quantifies the uncertainty therein to assist with model interpretability.

\end{abstract}

\section{Introduction}

The uses of machine learning (ML) have been attracting attention for various applications within the fluid dynamics community. 
In particular, ML approaches hold great potential for extracting complex nonlinear relations embedded in fluid flow data \cite{BNK2020,BEF2019,Kutz2017,FFT2020,BHT2020}. For example, the successes of ML have been observed in several investigations into the development of closure models for large eddy simulation (LES) \cite{maulik2019sub,gamahara2017searching,maulik2017neural,vollant2017subgrid} and Reynolds-Averaged Navier--Stokes (RANS) simulation \cite{ling2015evaluation,parish2016paradigm,ling2016reynolds,wu2018physics,matai2019zonal,zhu2019machine} where DNS or experimental data have been used to improve conventional algebraic or differential equation based closures \cite{DIX2019}. {On many occasions, physics-informed turbulence closure models have proven superior for canonical flows when compared to classical turbulence models with their restrictive hypotheses. However, the generalization of such models for broader applicability and their uncertainty quantification is only recently beginning to be explored \cite{geneva2019quantifying,geneva2020multi}.} Another promising avenue for ML is for addressing the challenges of conventional reduced order models (ROMs) \cite{THBSDBDY2020,THU2020}. Recent literature has demonstrated the capabilities of time-series methods from ML \cite{vlachas2018data,ahmed2019long,maulik2019time,mohan2019compressed,mohan2018deep,wang2019recurrent} for reduced-space temporal dynamics prediction as well as nonlinear subspace identification using image processing techniques \cite{lusch2018deep,gonzalez2018learning,xu2019multi,maulik2020reduced,cheng2020advanced,HFMF2019}. {A core motivation for exploring the use of ML methods for ROMs stems from the fact that ``intrusive''-ROMs depend on the resolution of a quadratic nonlinearity in POD-space \cite{rowley2017model} which is particularly computationally challenging in high-dimensional systems dominated by advection-type behavior. This leads to diminishing returns for equation-based ROMs when a large number of POD modes need to be resolved. There is a large volume of literature devoted to the stabilization (or closure) of intrusive-ROMs when there is improper spectral support to resolve all frequencies \cite{borggaard2011artificial,xie2018data,rezaian2020hybrid,san2015principal}. ML methods have also been used to learn these stabilizations to allow for hybrid modeling tasks where intrusive-ROMs are corrected by data \cite{san2018extreme,sheriffdeen2019accelerating}. This allows the ROM-user to balance the growing computational complexity of modal retention with data-driven corrections. Another limitation of POD-based ROMs comes into play when there is incomplete knowledge of the governing equations. This is encountered frequently when it is desirable to build emulators from experimental data. While less invasive ML-methods may still be employed to act as corrections to equation-based models, there is considerable interested in directly building an efficient and accurate emulation framework using the data alone. These ROMs are often termed \emph{non-intrusive} and have shown promising results for several applications where there may be a lack of underlying knowledge about the governing equations and data sets may suffer from incomplete observations \cite{kutz2016dynamic,kutz2016multiresolution,san2018neural,maulik2019time}}.

In addition to turbulence closures and ROMs, we have also seen successful applications for ML in fluid data estimation and reconstruction problems \cite{FNKF2019,erichson2019,FFT2019a,FFT2019b,OSM2019,LTHL2020,FFT2020b}. These efforts have revealed that ML, due to its inherent {ability to capture} nonlinearity \cite{Milano2002,MFF2019,FNF2020}, is well-suited to spatial flow data reconstruction and outperforms linear-reconstruction methods such as Gappy POD \cite{BDW2004} and linear stochastic estimation \cite{SH2017}. More recently, these investigated methods have also been applied to PIV data \cite{MFF2020,CZXG2019,DHLK2019} and understanding an interpretable relationship in between the predicted results and input data by focusing on vortical motions \cite{KL2020}.

Despite the aforementioned efforts, there are limited studies on interpretable ML methods, particularly for the widely used neural networks. This is partly due to the fact that most ML methods are utilized as ``black-boxes,'' which make point predictions. A maximum likelihood estimation of the loss function (usually a mean-squared-error or cross-entropy cost function) via gradient descent methods is usually executed for training, resulting in any trained neural network predicting the target values in a deterministic fashion. 

{{An alternative approach to obtain probabilistic outputs is to use Bayesian inference \cite{box2011bayesian}. This approach relies on interpreting each trainable parameter of the network to be a random variable that may be sampled from (for each prediction). This lends to an output that can be characterized with a probability density function. Such Bayesian Neural networks (for examples, see reviews by \cite{neal2012bayesian, MOHEBALI2020347, gal2016uncertainty, 2020arXiv200612024G} have been studied with various estimation schemes and applications. However, a Bayesian inference of the model parameters of the neural network is often prohibitively expensive in deeper architectures. Therefore several approximations using variational inference \cite{hinton1993, blundell2015, graves2011practical, 2013arXiv1312.6114K}, Monte-Carlo dropouts \cite{gal2015dropout, srivastava2014dropout}, Gaussian process approximations \cite{lu2013gp, damianou2019gp}{, kernel analog forecasting \cite{alexander2020operator}} and maximum \emph{a-posteriori} estimation \cite{santini95} are utilized to find the posterior of the weights during training. Most importantly, in the absence of such devices, a simple neural network result does not account for reliable uncertainty quantification, model selection or convergence requirements. The view of reliability for predictions is further crucial for more practical applications, which require transparency and accountability. }}

In the present paper, we {utilize a class of neural networks which assume that the predictions (in this case the observable) are generated from a distribution instead of providing just the best estimate. A broad class of such frameworks are referred to as mixture density networks (MDNs)  \cite{bishop1994mixture}. Several variants of this method \cite{ormoneit1996improved, yang2017hierarchical, makansi2019overcoming} have been developed over years and many have been successfully applied in various scientific applications (for instance, in \cite{2017arXiv171110746M, 2018A&A...609A.111D, 2020MNRAS.495.3087L}). The probabilistic neural network (PNN) we incorporate is a special case of MDN when the estimate is a single Gaussian distribution, instead of a mixture of Gaussians.} 

We then consider its applicability for several canonical problems of fluid dynamics. While the nature of the distribution is user-defined {and may be interpreted as an addition hyperparameter}, we utilize a unimodal Gaussian distribution. An attractive property of these networks is that a notion of uncertainty quantification is built into their architectures and any prediction for an observable is accompanied by an estimate for the corresponding uncertainty. This is obtained by assuming that the network predicts the mean and the variance of the posterior density function given the inputs it sees. The target is then compared to the mean of this density to compute a negative log-likelihood based distance metric (which is minimized).

The major highlights of this investigation are summarized as:
\begin{enumerate}
    \item We propose the use of a probabilistic neural network architecture for efficiently embedding uncertainty quantification into data-driven tasks relevant to fluid flows. 
    \item We execute several tests to assess the strengths of the probabilistic framework for model order reduction, forecasting and spatial field recovery across different applications.
    \item We also assess the robustness of the network architecture in the presence of noisy data and incomplete information.
\end{enumerate}

The present paper is organized as follows. The probabilistic neural network is introduced in Section 2. In Section 3, we apply the framework to representative problem settings for ROM and spatial fluid data recovery. We offer concluding remarks and future perspectives in Section 4.

\section{Probabilistic neural networks}

\begin{figure}
    \centering
    \includegraphics[width=0.8\textwidth]{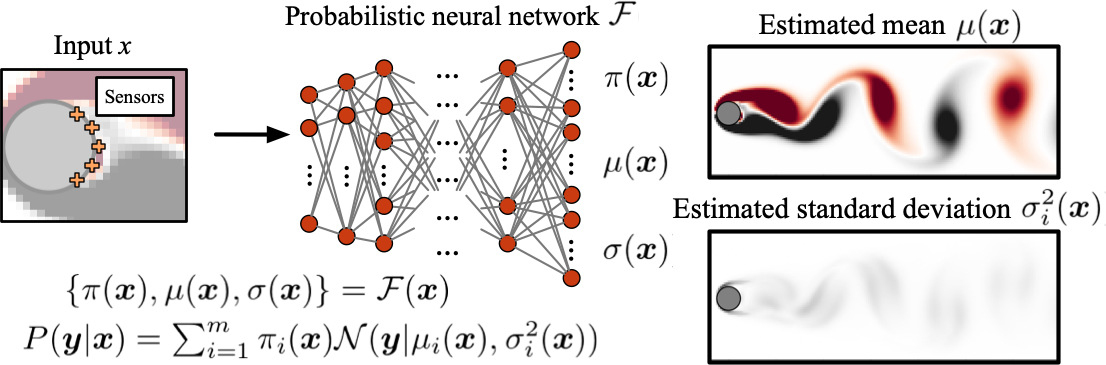}
    \caption{A representative schematic of the probabilistic neural network. We consider the flow reconstruction from sensor measurements of a two-dimensional cylinder wake (section \ref{sec:tcw}) as an example problem.}
    \label{fig:pnn}
\end{figure}

{Conventional neural networks approximate an arbitrary mapping from inputs to outputs using a number of trainable parameters called weights. However such networks either provide discrete classes as outputs (for classification problems) or continuous outputs (for regression). Generally a full description of the output, i.e., an output estimated as a probability distribution function (PDF) for given inputs is absent is such estimations. 
MDNs, initially proposed by \cite{bishop1994mixture} aim to approximate this conditional probability distribution, where the PDFs of the output, conditioned on the inputs are learned. The PDF is assumed to be a Gaussian mixture  distribution \cite{1988mmia.book.....M}. This output is effectively a posterior estimate that quantifies the uncertainty of the estimation a previously unseen data point.}

{MDNs are not the only network architecture to provide uncertainty estimates. Bayesian networks remain an active field of research \cite{mackay1995, gal2016uncertainty, neal2012bayesian, 2020arXiv200202405W}, with focus on scalabilty and interpretability of the posterior estimate. The probabilistic neural network framework we utilize is a special case of MDNs where the conditional probability estimate is assumed to a Gaussian. While MDNs with a large number of mixture components may approximate any arbitrary posterior estimate, it is achieved at a cost of computational expense and a larger requirement of training data.}  

\subsection{Network framework}

Predictions provided by neural networks for inputs $\bm x$ (such as the sensor measurements in figure \ref{fig:pnn}) generally are determined from the minimization of a loss function as a function of weights $\mathcal{E}({\bm w})$ (e.g., the mean squared error). Such calculations between the truths ${\bm y}_t$ (such as the whole field in figure \ref{fig:pnn}) and the predicted values from the network ${\bm y}_p({\bm x})$ lack a probability distribution of $p({\bm y}_p|{\bm x})$. Hence, the error bars on the estimates are typically absent in dense neural network outputs. To quantify uncertainty of the neural network estimates, the multi-dimensional surface of the loss function $\mathcal{E}({\bm w})$ has to be explored in addition to finding the global minimum. 

To obtain the posterior of estimates $p({\bm y}_p|{\bm x})$, one may sample the network weights, thus providing a fully Bayesian inference on the parameter estimates. However, this method is computationally expensive unless the networks are shallow. For this reason, obtaining the probability distribution functions of the predictions in a computationally feasible manner is a challenging problem. 
To address this issue, we approach this problem by avoiding sampling altogether. 
Instead of mapping the neural network from inputs $\mathcal{F}: {\bm x} \rightarrow {\bm y}({\bm x})$, one could define the mapping as $\mathcal{F}: {\bm x} \rightarrow (\mu, \sigma)$ where the mean $\mu$ and standard deviation $\sigma$ parametrize a Gaussian probability distribution function $\mathcal{N}(\mu,\sigma)$. 
This approach ensures that the outputs $p({\bm y}_p|{\bm x}) = \mathcal{N}(\mu({\bm x}),\sigma({\bm x}))$ are accompanied by uncertainty estimates assuming the errors are Gaussian. 

For more complex probability distributions that depart from the Gaussian distribution, one may parametrize the mapping distribution accordingly. One straightforward extension is the Gaussian mixture model with the mapping  $\mathcal{F}: {\bm x} \rightarrow (\pi_1, \mu_1, \sigma_1, \pi_2, \mu_2, \sigma_2, ...,  \pi_N, \mu_N, \sigma_N)$, where $\pi_i$ is the mixing probability for each Gaussian component satisfying the condition $\sum_{i=1}^{m}\pi_i = 1$. The distribution function in this model is a linear combination of several Gaussian components given by,
\begin{equation}\label{eq:GMM}
    p({\bm y} | {\bm x}) = \sum_{i=1}^{m}\pi_i ({\bm x}) \mathcal{N}(\mu_i ({\bm x}),\sigma_i ({\bm x})).
\end{equation}


The value of $m$ is generally pre-specified based on the expectation of posterior distribution. For example, if the output distribution is expected to be bimodal, one may select $m=2$. The main advantage of this Gaussian mixture modeling, however, is that extremely complicated complex distributions can be approximated as a mixture of Gaussians with a large number of mixing components (large $N$). 
Alternatively, the number of mixing components can also be independently optimized along with learning rate, decay rate and other hyper-parameters of the training scheme. It has to be noted that the training of MDN itself is supervised. However, mixture probabilities that are obtained from a fully trained network corresponds to different clusters in the data that is learned in an unsupervised fashion.  

\begin{algorithm}[H]
\label{algo:pnn}
\caption{Training procedure for probabilistic neural network $\cal F$}
\begin{algorithmic}[1]
\STATE ${\bm w} \leftarrow $ Initialize network parameters
\WHILE{{$\mathcal{E}$ is not converged}}
\STATE Update probabilistic neural network
\STATE $x \leftarrow  $Random mini-batch from data set
\FOR{$x_k$ in $x$} 
\STATE Compute Gaussian mixture parameters: $(\pi_k, \mu_k, \sigma_k)$ $\leftarrow$ ${\cal F}(x_k; {\bm w})$ 
\STATE Compute predictive distribution: $p(y_{k,p}|x_k) \leftarrow (\pi_k, \mu_k, \sigma_k)$ 
\ENDFOR
\STATE  Compute negative log-likelihood: $\mathcal{E} \leftarrow -\log \mathcal{L}$ 
\STATE Update network parameters: ${\bm w} \leftarrow { Adam}(\nabla_{\bm w} \mathcal{E}, {\bm w})$ 
\ENDWHILE
\end{algorithmic}
\end{algorithm}

Since our PNN provides a distribution of the estimation $p({\bm y}_p|{\bm x})$ instead of a point prediction of ${\bm y}_p$, the loss function has to chosen accordingly to utilize the full distribution of the prediction. This results in a crucial improvement conventional neural networks. We choose to maximize the average likelihood of the training data, which captures the full information about the entropy between the distribution of the training data ${\bm y}_t$ and the corresponding prediction ${\bm y}_p$. Hence in practice, our model ${\mathcal F}({\bm x};{\bm w})$ is trained to obtain the optimized weights $\bm w$ by minimizing the error function $\mathcal{E}$ given in terms of average log-likelihood $\mathcal{L}$ such that
\begin{align}\label{eq:negloglike}
\begin{gathered}
{\bm w}={\rm argmin}_{\bm w}[\mathcal{E}],\quad \text{where} \quad \mathcal{E} \equiv -\log \mathcal{L} = -\sum_{k=1}^{K} p({y}_{k,p}|{x_k}) \log p({y}_{k,t}),
\end{gathered}
\end{align}
with $k$ indicating each data point in the training data and $K$ denotes a number of training samples. The term $p({y}_{p,k}|{x_k})$ in the error function is evaluated for each data point using the output of the network given in equation \ref{eq:GMM}. 

{Cross-entropy is a common error function used for classification problems using neural networks. In a continuous limit (i.e, in regression problems like ours), the cross-entropy reduces to mean square error between the outputs of the neural network and the targets. On the other hand, negative log-likelihood shown in equation \ref{eq:negloglike} also reduces to mean square error when mean predictions alone are considered (i.e., in point-predictions instead of probability distribution outputs). In other words, log-likelihood and cross-entropy are analogous cost functions in regression and classification problems, respectively \cite{nielsen2015neural}. Thus the likelihood maximizing model is equivalent to minimizing the cross entropy $H(p({\bm y}_p|{\bm x}), p({\bm y}_t) )$. Hence our framework incorporates a generic loss prescription that can be applied to a wide variety of problems and specifies a representative uncertainty in the machine-learned estimates.}

Our implementation of a PNN is shown for one of the applications (section 3.1) in figure \ref{fig:pnn}. In this example, the inputs ${\bm x}\in{\mathbb R}^5$ to the network model are the sensor measurements. These are mapped to the targets ${\bm y}_p$ whose truth values are a whole wake field of two-dimensional cylinder ${\bm y}_t$ of dimension 13440.  The network for the example of figure \ref{fig:pnn} is a fully connected dense network with 8 layers and number of neurons per layer {having} $5\rightarrow 64 \rightarrow 128 \rightarrow 256 \rightarrow 512 \rightarrow 1024 \rightarrow 2048 \rightarrow 13440\times 3$. 
The last layer [$13440\times 3$] corresponds to $(\pi_1, \mu_1, \sigma_1)$ for each grid point that parametrizes the predictive conditional distribution $p({\bm y}_p|{\bm x})$ for the whole wake field, and the error function is calculated using equation \ref{eq:negloglike}. We remind the reader that we use only one Gaussian center in this investigation and focus on one value each of mean and variance ($\mu_1$ and $\sigma_1$), which implies $\pi_1 = 1$ corresponding to each target. We utilize the same number of Gaussian centers (i.e., solely one) for all our assessments hereafter. As an example, if 3 Gaussian centers were utilized, the final layer of the above map would have dimensions of [$13440\times 3 \times 3$] corresponding to $(\pi_i, \mu_i, \sigma_i)$ for $i=1,2,3$ and $\sum_{i=1}^3 \pi_i = 1$.

For updating the weights $\bm w$, we use the Adam optimizer \cite{Kingma2014} and the training procedure of the PNN is summarized in algorithm 1. We note that we do not attempt to optimize the parameters of the network architecture for each problem setting in the present study. Our primary objective here is to demonstrate the applicability of the probabilistic model to quantify uncertainty of machine-learned estimations for canonical fluid flow problems. One may consider the use of theoretical optimization methods such as Hyperopt \cite{BYC2013} and Bayesian optimization \cite{maulik2019time,BCF2009} to further enhance the accuracy of the estimation can be improved. A sample code for the present model is available online \cite{MFRFT2020code}. { The code for the present study has been written in-house by utilizing TensorFlow 2.2.0 on Python 3.7. The present computation for the probabilistic neural network is performed on Nvidia Tesla V100 graphics processing unit (GPU).} 

\section{Results}

\begin{figure}
    \centering
    \includegraphics[width=0.95\textwidth]{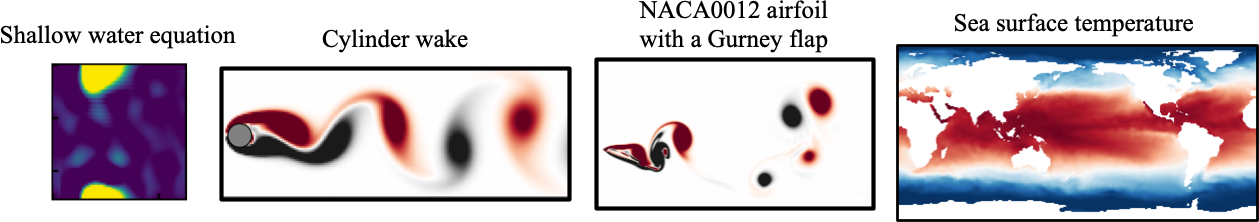}
    \caption{The different canonical test-cases considered in the present work.}
    \label{fig:dataset}
\end{figure}

In this section, we demonstrate the capabilities of the PNN introduced in the previous section through a variety of experiments. These experiments are:
\begin{itemize}
    \item The prediction of POD coefficient evolution given initial conditions (sec. \ref{sec:3-1}), 
    \item The instantaneous estimation of POD coefficients from local sensor measurements (sec. \ref{sec:3-2}).
\end{itemize}
Next, we consider the applications to spatial fluid data recovery given by:
\begin{itemize}
    \item The estimation of sensor measurements from other sensor placements,
    \item The estimation of whole flow fields from local sensor measurements.
\end{itemize}
for three example problems:
\begin{itemize}
    \item A two-dimensional cylinder wake at ${\rm Re}_D=100$ (sec. \ref{sec:tcw}),
    \item A two-dimensional wake of NACA0012 airfoil with a Gurney flap at ${\rm Re}_c=1000$ (sec. \ref{sec:ngf}),
    \item The NOAA optimum interpolation sea surface temperature data set \cite{noaa} (sec. \ref{sec:noaasst}).
\end{itemize}
The broad spread of potential applications can be observed through flow fields from our chosen data sets as seen in figure \ref{fig:dataset}.

\subsection{Parameteric surrogates}
\label{sec:3-1}

In the following, we review the POD technique for the construction of a reduced basis \cite{THBSDBDY2020,TBDRCMSGTU2017,kosambi1943statistics,berkooz1993proper} for our observed variable. This is for the express purpose of easily implementing a data-driven recovery of the system evolution in reduced-space. The POD procedure is tasked with identifying a space
\begin{linenomath*}
\begin{align}
\mathbf{X}^{f}=\operatorname{span}\left\{\boldsymbol{\vartheta}^{1}, \dots, \boldsymbol{\vartheta}^{f}\right\},
\end{align}
\end{linenomath*}
which approximates snapshots $\boldsymbol{\vartheta}^i$ optimally with respect to the $L_2$ norm. The process of $\boldsymbol{\vartheta}$ generation commences with the collection of snapshots in the \emph{snapshot matrix}
\begin{linenomath*}
\begin{align}
\mathbf{S} = [\begin{array}{c|c|c|c}{\hat{\bm{q}}^{1}_h} & {\hat{\bm{q}}^{2}_h} & {\cdots} & {\hat{\bm{q}}^{N_{s}}_h}\end{array}] \in \mathbb{R}^{N_{h} \times N_{s}},
\end{align}
\end{linenomath*}
where $N_s$ is the number of snapshots, and $\hat{\bm{q}}^i_h : \mathcal{T} \times \mathcal{P} \rightarrow \mathbb{R}^{N_h}$ corresponds to an individual snapshot in time of the discrete solution domain with the mean value removed, i.e.,
\begin{linenomath*}
\begin{align}
\begin{gathered}
\hat{\bm{q}}^i_h = \bm{q}^i_h - \bm{\bar{q}}_h, ~~~\bm{\bar{q}}_h = \frac{1}{N_s} \sum_{i=1}^{N_s} \bm{q}^i_h.
\end{gathered}
\end{align}
\end{linenomath*}
with $\overline{\bm{q}}_h : \mathcal{P} \rightarrow \mathbb{R}^{N_h}$ being the time-averaged solution field. Our POD bases can then be extracted efficiently through the method of snapshots where we solve the eigenvalue problem on the correlation matrix $\mathbf{C} = \mathbf{S}^T \mathbf{S} \in \mathbb{R}^{N_s \times N_s}$. Then
\begin{linenomath*}
\begin{align}
\begin{gathered}
\mathbf{C} \mathbf{W} = \mathbf{W} \mathbf{\Lambda},
\end{gathered}
\end{align}
\end{linenomath*}
where $\mathbf{\Lambda} = \operatorname{diag}\left\{\lambda_{1}, \lambda_{2}, \cdots, \lambda_{N_{s}}\right\} \in \mathbb{R}^{N_{s} \times N_{s}}$ is the diagonal matrix of eigenvalues and $\mathbf{W} \in \mathbb{R}^{N_{s} \times N_{s}}$ is the eigenvector matrix. Our POD basis matrix can then be obtained by
\begin{linenomath*}
\begin{align}
\begin{gathered}
\boldsymbol{\vartheta} = \mathbf{S} \mathbf{W} \in \mathbb{R}^{N_h \times N_s}.
\end{gathered}
\end{align}
\end{linenomath*}
In practice a reduced basis $\boldsymbol{\psi} \in \mathbb{R}^{N_h \times N_r}$ is built by choosing the first $N_r$ columns of $\boldsymbol{\vartheta}$ for the purpose of efficient ROMs, where $N_r \ll N_s$. This reduced basis spans a space given by
\begin{linenomath*}
\begin{align}
\mathbf{X}^{r}=\operatorname{span}\left\{\boldsymbol{\psi}^{1}, \dots, \boldsymbol{\psi}^{N_r}\right\}.
\end{align}
\end{linenomath*}
The coefficients of this reduced basis (which capture the underlying temporal effects) may be extracted as
\begin{linenomath*}
\begin{align}
\begin{gathered}
\mathbf{A} = \boldsymbol{\psi}^{T} \mathbf{S} \in \mathbb{R}^{N_r \times N_s}.
\end{gathered}
\end{align}
\end{linenomath*}
The POD approximation of our solution is then obtained via
\begin{linenomath*}
\begin{align}
\hat{\mathbf{S}} =  [\begin{array}{c|c|c|c}{\tilde{\bm{q}}^{1}_h} & {\tilde{\bm{q}}^{2}_h} & {\cdots} & {\tilde{\bm{q}}^{N_{s}}_h}\end{array}] \approx \boldsymbol{\psi} \mathbf{A} \in \mathbb{R}^{N_h \times N_s},
\end{align}
\end{linenomath*}
where $\tilde{\bm{q}}_h^i : \mathcal{T} \times \mathcal{P} \rightarrow \mathbb{R}^{N_h}$ corresponds to the POD approximation to $\hat{\mathbf{q}}_h^i$. The optimal nature of reconstruction may be understood by defining the relative projection error
\begin{linenomath*}
\begin{align}
\frac{\sum_{i=1}^{N_{s}}\left\|\hat{\bm{q}}^i_h-\tilde{\bm{q}}^i_h \right\|_{\mathbb{R}^{N_{h}}}^{2}}{\sum_{i=1}^{N_{s}}\left\|\hat{\bm{q}}^i_h\right\|_{\mathbb{R}^{N_{h}}}^{2}}=\frac{\sum_{i=N_r+1}^{N_{s}} \lambda_{i}^{2}}{\sum_{i=1}^{N_{s}} \lambda_{i}^{2}},
\end{align}
\end{linenomath*}
which exhibits that with increasing retention of POD bases, increasing reconstruction accuracy may be obtained. Our first application of PNNs will be tasked with predicting $\tilde{\bm{q}}_h^i$ given information about parameters that control the evolution of the system. 

\subsubsection{The inviscid shallow water equations}
\label{SWE_ROM}

Our first example considers the two-dimensional inviscid shallow water equations which are a prototypical system for geophysical flows. The governing equations are hyperbolic in nature and are
\begin{align}
    \label{Eq_SWE}
    \begin{gathered}
    \frac{\partial(\rho \eta)}{\partial t}+\frac{\partial(\rho \eta u)}{\partial x}+\frac{\partial(\rho \eta v)}{\partial y} =0, \\
    \frac{\partial(\rho \eta u)}{\partial t}+\frac{\partial}{\partial x}\left(\rho \eta u^{2}+\frac{1}{2} \rho g \eta^{2}\right)+\frac{\partial(\rho \eta u v)}{\partial y} = 0, \\
    \frac{\partial(\rho \eta v)}{\partial t}+\frac{\partial(\rho \eta u v)}{\partial x}+\frac{\partial}{\partial y}\left(\rho \eta v^{2}+\frac{1}{2} \rho g \eta^{2}\right) = 0.
    \end{gathered}
\end{align}
In the above set of equations, $\eta$ corresponds to the total fluid column height, and $(u,v)$ is the horizontal flow velocity of the fluid averaged across the vertical column. Furthermore, $g$ is gravitational acceleration, and $\rho$ is the fluid density fixed to be unity. The first equation represents conservation of mass and the remaining two denote the conservation of momentum. For simplicity, we denote $\bm{q} = [\rho \eta, \rho \eta u, \rho \eta v]^T$ and we use $\eta$ and $\rho \eta$ interchangeably hereafter. We also point out that there is no linear diffusion term in the above shallow-water equations. This implies that our system evolution is solely advection-dominated with any dissipation occurring as a result of numerical viscosity alone. Our initial conditions are given by 
\begin{align}
\begin{gathered}
    \rho \eta (x,y,t=0) = 1+{\rm exp}\left[{-\left(\frac{(x-\bar{x})^2}{2(5e+4)^2} + \frac{(y-\bar{y})^2}{2(5e+4)^2}\right)}\right], \\
    \rho \eta u(x,y,t=0) = 0, ~~~\rho \eta v(x,y,t=0) = 0,
\end{gathered}
\end{align}
while our two-dimensional domain is a square with periodic boundary conditions. These initial conditions $\bm{r} = [\bar{x},\bar{y}]$ ($-0.5\leq \{\bar{x},\bar{y}\} \leq 0.5$) correspond to the parameter controlling the spatio-temporal evolution of the dynamical system. 

Our data generation process utilizes full-order solves of the above system of equations until $t=0.1$ with a time step of 0.001. Our full-order model uses a 4\textsuperscript{th}-order accurate Runge-Kutta integration scheme and a fifth-order accurate weighted essentially non-oscillatory scheme (WENO) \cite{liu1994weighted} for computing state reconstructions at cell faces. The Rusanov Reimann solver is utilized for flux reconstruction after cell-face quantities are calculated. The reader is directed to \cite{hairer1991solving} for greater discussion of the temporal integration scheme and \cite{maulik2017resolution} for details on WENO and the Riemann solver implementation in two-dimensional problems. 

\begin{figure}[t]
    \centering
    \includegraphics[width=0.95\textwidth]{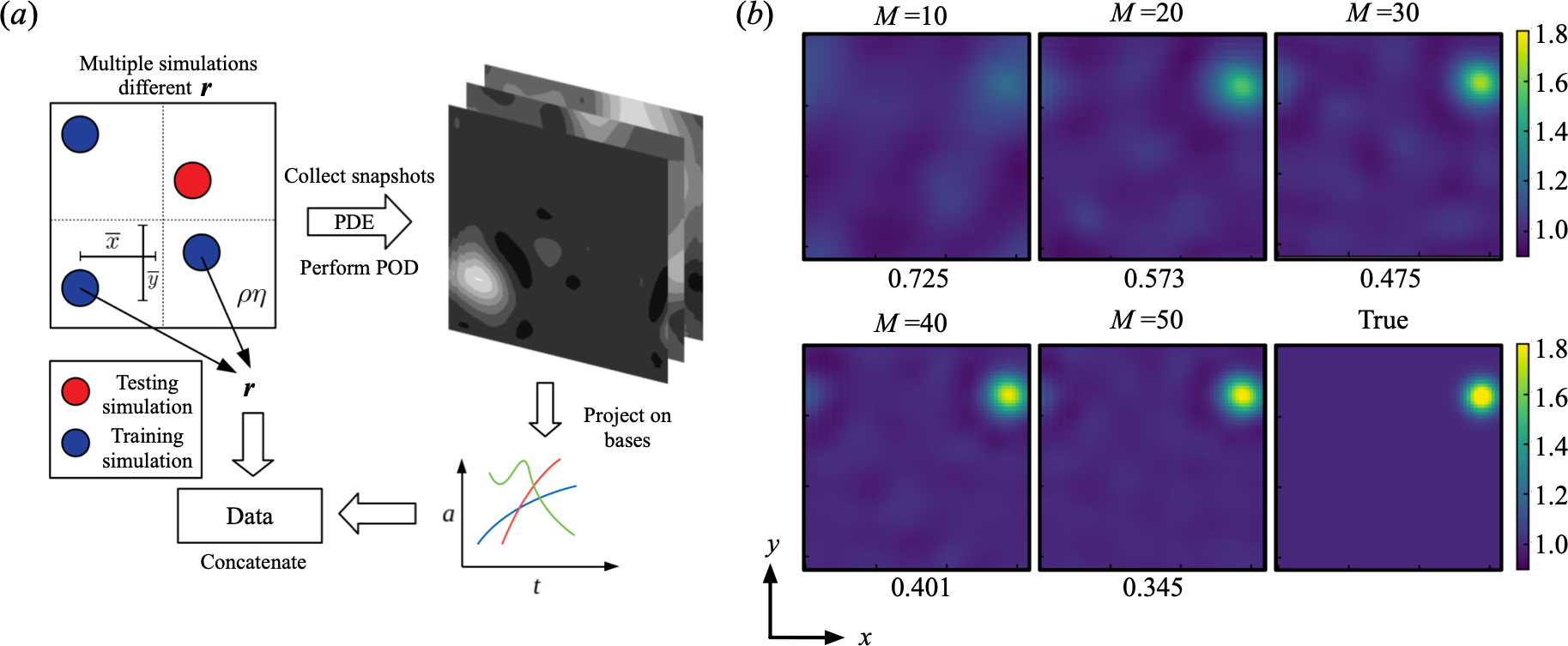}
    \caption{{ Information of the data set for the shallow water equation.} $(a)$ A schematic for the generation of training and testing data. Given inputs of $\bm{r}$ (i.e., the location of an initial Gaussian perturbation to $\eta$), the machine learning ROM is tasked with predicting the evolution of 40 POD coefficients over 10 snapshots in time. The prediction is one-shot. $(b)$ Convergence of reduced representations to true field with increasing $M$. The values underneath each field indicate {\color{black} the normalized $L_1$ error norm $\epsilon = |\eta_{\rm True}-\eta_{\rm POD}|/|\eta_{\rm True}-1|$}.}
    \label{fig:PCA_Schematic}
\end{figure}

The procedure for generating the training and testing data is illustrated in figure \ref{fig:PCA_Schematic}$(a)$.
A hundred different locations $\bm{r} = [\bar{x}, \bar{y}]$ are chosen for full-order simulations. We note that these 100 locations are chosen by Latin hypercube sampling. Out of the 100 simulations for which snapshots are obtained, 90 simulations are set aside for the purpose of training machine learning frameworks and for POD basis generation. 
Each simulation is sampled 10 times temporally to obtain a total of 900 training snapshots for POD basis generation.
Coefficients obtained by projecting these training simulations onto the POD bases are configured to be the target data for training the machine learning frameworks. 
For accurately capturing the evolution of the perturbation, we choose $M=40$ as the number of POD coefficients with {\color{black}the normalized $L_1$ error norm $\epsilon = |\eta_{\rm True}-\eta_{\rm POD}|/|\eta_{\rm True}-1|$ of 0.401}, as shown in figure \ref{fig:PCA_Schematic}$(b)$. 
All our assessments for the machine learning frameworks are performed on snapshots obtained from the 10 simulations kept aside for the purpose of testing. 
Also, our assessments are performed solely for the $\eta$ flow field. We emphasize on this point, since we assume that the other key variables $\rho \eta u$ and $\rho \eta v$ are not available for observation. This aligns with practical applications for fluid dynamics where constraints of cost or safety make it impractical to observe all the physics of a system. We would also draw attention to the fact that the initial condition of the entire system is conditioned on a Gaussian excitation to $\rho \eta$ alone. Since the other conserved variables are never observed for generating training data, the framework cannot be expected to work if there are unseen perturbations to their initial conditions.

We first assess the ability of the proposed PNN for predicting the evolution of the shallow-water equation dynamics given an input of ${\bm q}=[\bar{x}, \bar{y}]$ alone. In this task, we have a two-dimensional input space $\bm q$ and a 400-dimensional output space ${\bm a}_t=[{\bm a}^1,{\bm a}^2,...,{\bm a}^{10}]$ corresponding to 40 spatial POD coefficients ${\bm a}^{\iota}=[a^{\iota}_1,a^{\iota}_2,...,a^{\iota}_{40}]$ over 10 time steps so as to account for the entire trajectory of the $\eta$ dynamics. Following equation \ref{eq:GMM} and figure \ref{fig:pnn}, the problem setting here can be formulated as 
\begin{align}
\begin{gathered}
\{\pi({\bm q}),\mu({\bm q}),\sigma({\bm q})\}={\cal F}({\bm q}),~~~p({\bm a}_t | {\bm q}) = \sum_{i=1}^{m}\pi_i ({\bm q}) \mathcal{N}(\mu_i ({\bm q}),\sigma_i ({\bm q})).
\end{gathered}
\end{align}
{For the purpose of effective training, all POD coefficients are scaled between 0 and 1}. A schematic for this task and its assessment is shown in figure \ref{fig:PCA_Schematic}$(a)$.

\begin{figure}
    \centering
    \includegraphics[width=0.75\textwidth]{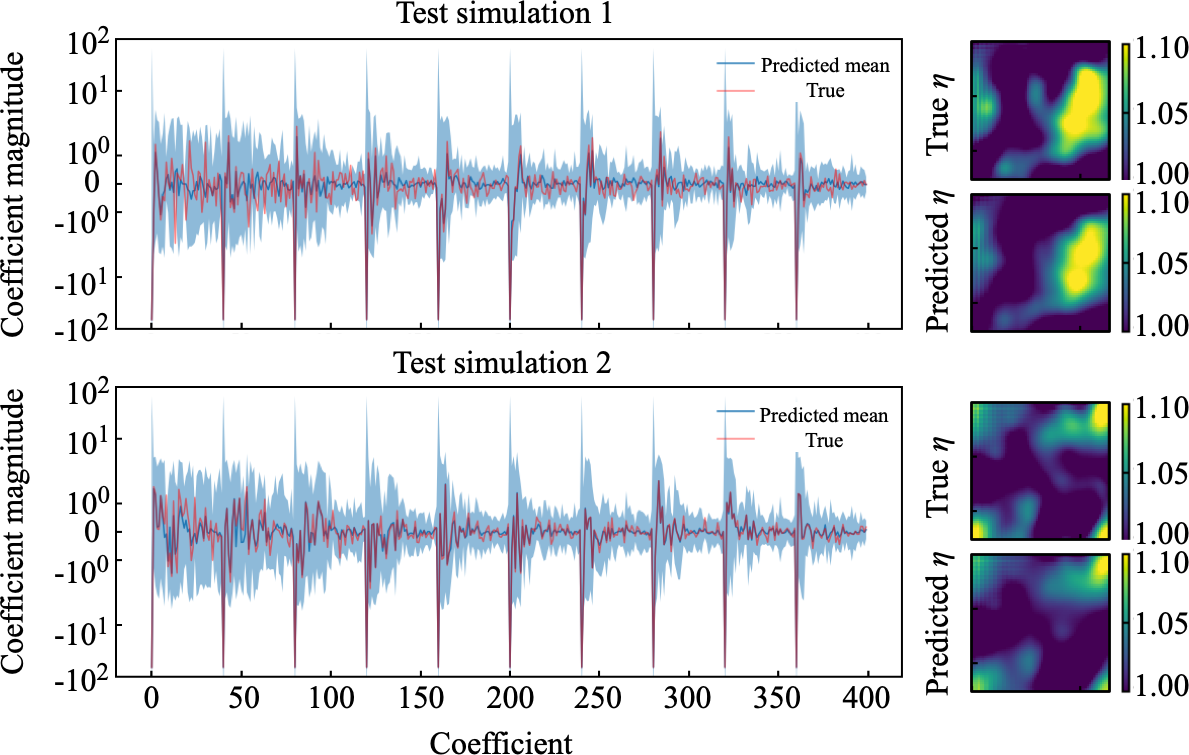}
    \caption{Coefficient estimations of two representative test simulation by the present framework {with shallow water equation} showing error bars for two standard deviations around the mean as parameterized by a Gaussian distribution. Final time field reconstructions using estimated POD coefficient means are shown in the right side of each coefficient evolution. True indicates the reference field with 40 POD modes.}
    \label{fig:PNN_Final_Field}
\end{figure}

The coefficient estimations from the present framework for two representative simulations are shown in figure \ref{fig:PNN_Final_Field}. The 400 coefficients are presented sequentially with 40 coefficients representing the information needed to reconstruct the flow-field at one time instant. The coefficients are also accompanied by confidence intervals spanning two standard deviations on either side of the predicted means and represent the uncertainty quantification mechanism built into the probabilistic neural network. The dissipation of the coherent structures at later snapshots also leads to a reduced prediction of uncertainty by the framework. Whereas at earlier snapshots, much larger uncertainty is observed due to the present of coherent distortions of the solution field. The corresponding field reconstructions using the mean coefficients can be seen in the right side of each coefficient evolution where qualitative agreement with the true simulations is clearly observable. 
{\color{black}We obtain the normalized $L_1$ error norm of 0.843 for the reconstructed field of test simulations.}

{At this juncture, we provide some remarks on the nature of the system that is emulated. The initial and boundary conditions for this particular shallow-water equation experiment represent a tightly-controlled traveling wave problem that is translation invariant. Different realizations of the initial condition lead to translationally shifted trajectories. We also note the presence of mirror symmetries with respect to $x=\bar{x}$ and $y=\bar{y}$ coupled with a rotational symmetry of $\pi/2$ radians about the origin. \textcolor{black}{We caution the reader that the current ROM strategy is devoid of any symmetry preserving mechanisms. In fact, the projection of the true field onto the 40 dominant POD modes breaks the aforementioned symmetries and thus leads to the lack of symmetry preservation within the training data itself.} However, our motivation for a first assessment of our PNN on this system stems from the well-known fact that ROMs obtained from POD-based compression are severely limited in their ability to forecast on these simple traveling-wave systems \cite{KFB2016,mendible2019dimensionality} and require special treatment with intrinsic knowledge of the flow dynamics.} Also, the surrogate model proposed here is conditioned on the initial flow field of $\rho \eta$ alone. Thus, this framework represents a promising approach for incomplete observations of geophysical dynamics where complete knowledge of physics is almost always impossible. We further this claim by actual experiments on remote sensing data sets later in this study.

\subsubsection{{NOAA sea surface temperature}}
\label{sec:3-1-2}

{To demonstrate the applicability of the PNN to practical applications, let us consider the NOAA sea surface temperature data set. The field data here has the spatial resolution of $360\times180$ based on a one degree grid and is obtained from satellite and ship observations without adequate knowledge of underlying governing equations. We use 20 years of data (1040 snapshots spanning 1981 to 2001) as the training data set, while the test data set is prepared from 874 snapshots spanning from year 2001 to 2018. This test setting is extrapolation in time but not physics, since the data set has influence of seasonal periodicity. The aforementioned problem setting follows the work of Callaham et al. \cite{CMB2019} who attempted to reconstruct fluid flow fields from local sensors using sparse representations. Following Callaham et al. \cite{CMB2019}, the input sensors for the baseline model are chosen randomly from the region of $50{^\circ}$ S to $50{^\circ}$ N.
}

\begin{figure}
    \centering
    \includegraphics[width=0.98\textwidth]{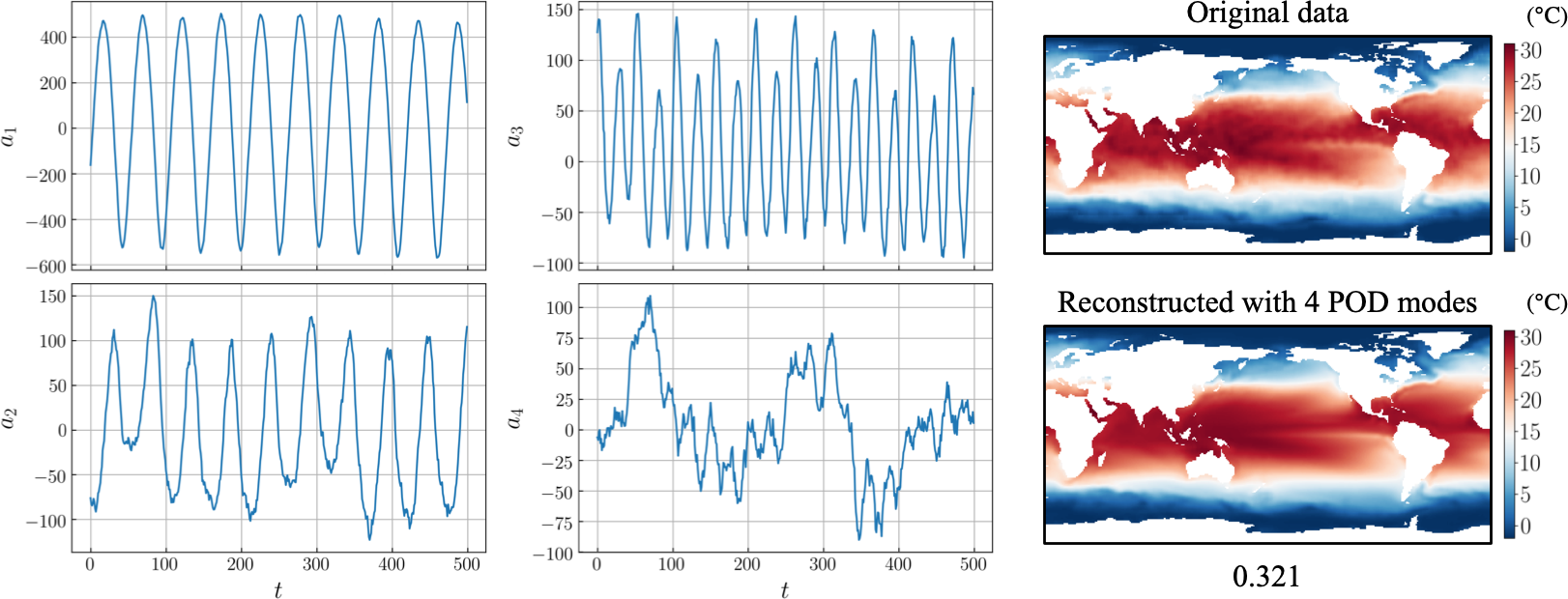}
    \caption{{POD coefficients and entire field of NOAA sea surface temperature. The value underneath the contour is the $L_2$ error norm $\epsilon_{\rm POD}=||T_{\rm ref}-T_{\rm POD}||_2/||T_{\rm ref}||_2$}.}
    \label{fig:noaa_rom_setup}
\end{figure}

{
For the task of constructing a parametric surrogate for forecasting the sea surface temperature, the PNN attempts to predict the temporal evolution of four POD coefficients ${\bm a}^{\iota}=[a^{\iota}_1,a^{\iota}_2, a^{\iota}_{3},a^{\iota}_{4}]$ over 100 weeks ${\bm a}_t=[{\bm a}^1,{\bm a}^2,...,{\bm a}^{100}]$ from the local sensor measurements on the first week snapshot ${\bm s}^1$, i.e., initial information. The problem setting is expressed as 
\begin{align}
\begin{gathered}
\{\pi({\bm s}^1),\mu({\bm s}^1),\sigma({\bm s}^1)\}={\cal F}({\bm s}^1),~~~p({\bm a}_t | {\bm s}^1) = \sum_{i=1}^{m}\pi_i ({\bm s}^1) \mathcal{N}(\mu_i ({\bm s}^1),\sigma_i ({\bm s}^1)).
\end{gathered}
\end{align}
As shown in figure \ref{fig:noaa_rom_setup}, the reconstructed field with 4 spatial POD modes shows qualitative agreement with the original data globally, while the seasonal patterns at the global scale can be captured with modes 1, 2, and 3 and stochastic fluctuations may be represented with mode 4 \cite{MELB2020}.
}

\begin{figure}
    \centering
    \includegraphics[width=0.98\textwidth]{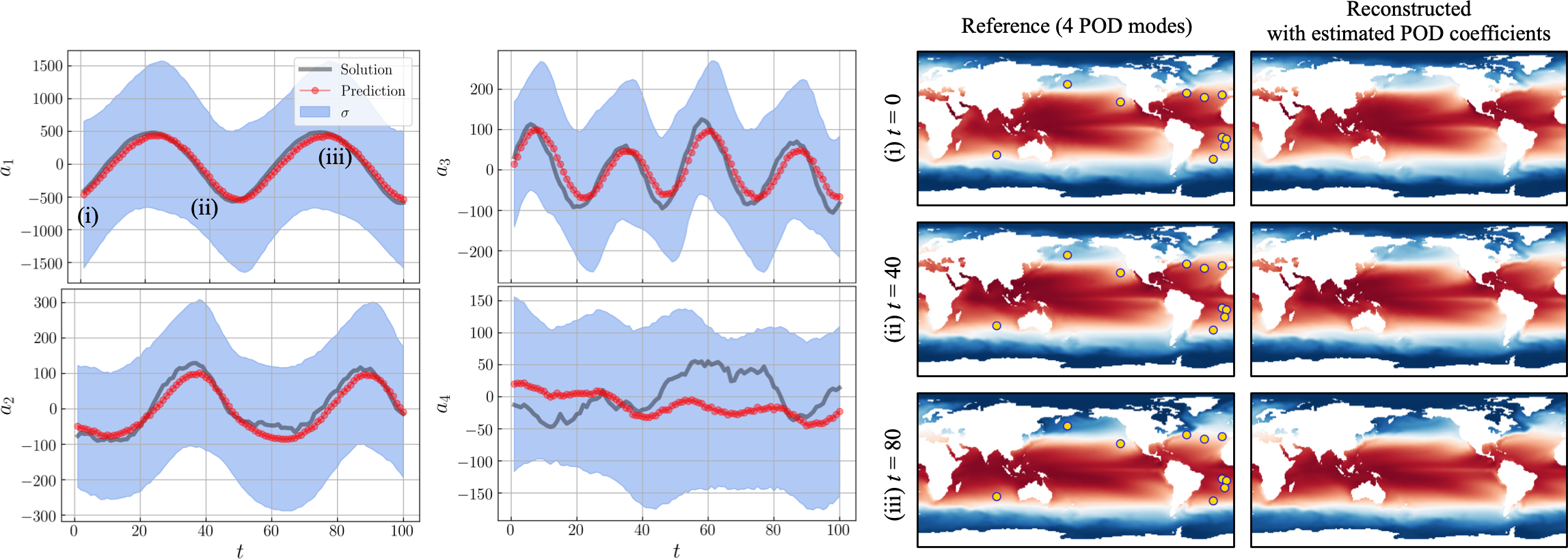}
    \caption{{Prediction of POD coefficient evolution from sensors at the first snapshot. Representative temporal evolution and the reconstructed field with POD eigenvectors at the first, 40th and 80th weeks are shown.}}
    \label{fig:noaa_rom_base}
\end{figure}

{The baseline results predicted from 10 input sensors are summarized in figure \ref{fig:noaa_rom_base}. The periodic trends represented by modes 1, 2, and 3 can be reasonably captured from only 10 sensor measurements on the first snapshot while telling us the confidence interval of its estimation. Note, however, that the predicted temporal evolution of mode 4 which represents more stochasticity than that of the first 3 modes exhibits further greater deviation compared to the reference. This result provides motivations to examine the influence on the number of input sensors, as discussed in the rest of this article. 
}

{

\begin{figure}
    \centering
    \includegraphics[width=1.0\textwidth]{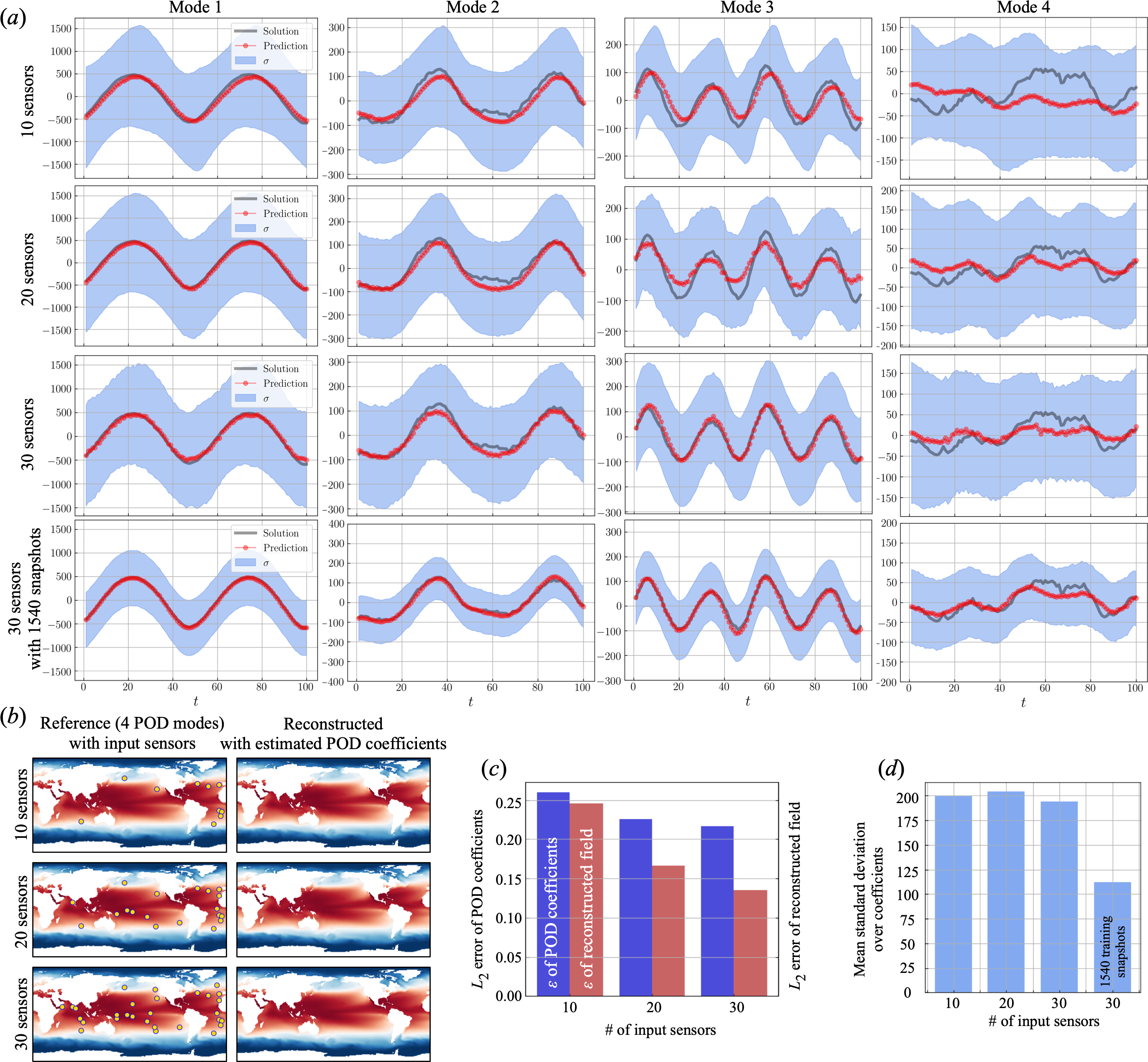}
    \caption{{Dependence of the POD coefficients prediction for sea surface temperature on the number of input sensors. {\color{black}$(a)$ Temporal evolution of 4 POD coefficients.} $(b)$ Reconstructed temperature field with POD eigenvectors. $(c)$ $L_2$ error norm of predicted POD coefficients and reconstructed field. {\color{black}$(d)$ Ensemble average of estimated standard deviation taking over all POD coefficients.}}}
    \label{fig:noaa_rom_inpsen}
\end{figure}

We investigate the dependence of the prediction accuracy on the number of input sensors, as shown in figure \ref{fig:noaa_rom_inpsen}. 
The curve starts to match with the reference data on increasing the number of input sensors, which can be particularly seen with mode 3 in figure \ref{fig:noaa_rom_inpsen}$(a)$. The similar trends can be confirmed in both figures \ref{fig:noaa_rom_inpsen}$(b)$ and $(c)$ which show the reconstructed fields and $L_2$ error norms.
Regarding the standard deviations presented in figure \ref{fig:noaa_rom_inpsen}$(d)$, the estimated uncertainties show no significant difference over the covered number of inputs{\color{black}, while decreasing by adding the training snapshots}. 
This is due to the fact that the confidence interval of the present framework is not a barometer for error {\color{black}but a measure of the quality of the training data}.
This point can also be found in the later examples.
}

\subsection{Field reconstructions through POD coefficient estimation { for the shallow water equations}}
\label{sec:3-2}

In this section, we demonstrate the capability of the proposed framework to reconstruct the state of the field by random sampling in the domain. Similar to the previous section, we task the probabilistic neural network in predicting the POD coefficients $\bm a$. Our inputs, however, are now given by sensor measurements $\bm s$ of the field in addition to a time stamp $q_t$ that indicates the progress to the final time of the evolution. Hence, the problem setting here can be expressed as 
\begin{align}
\begin{gathered}
\{\pi({[{\bm s},{q_t}]}),\mu([{\bm s},{q_t}]),\sigma([{\bm s},{q_t}])\}={\cal F}([{\bm s},{q_t}]),~~~p({\bm a} | [{\bm s},{q_t}]) = \sum_{i=1}^{m}\pi_i ([{\bm s},{q_t}]) \mathcal{N}(\mu_i ([{\bm s},{q_t}]),\sigma_i ([{\bm s},{q_t}])).
\end{gathered}
\end{align}
Note that this estimation is performed at the same instantaneous field between the input and output. We clarify that the random sensors that are measured for the purpose of field reconstruction are not perturbed by noise at the moment. {We also compare the performance of the PNN against the well-known Gappy POD technique for flow-field recovery}.


\begin{figure}
    \centering
    \includegraphics[width=0.95\textwidth]{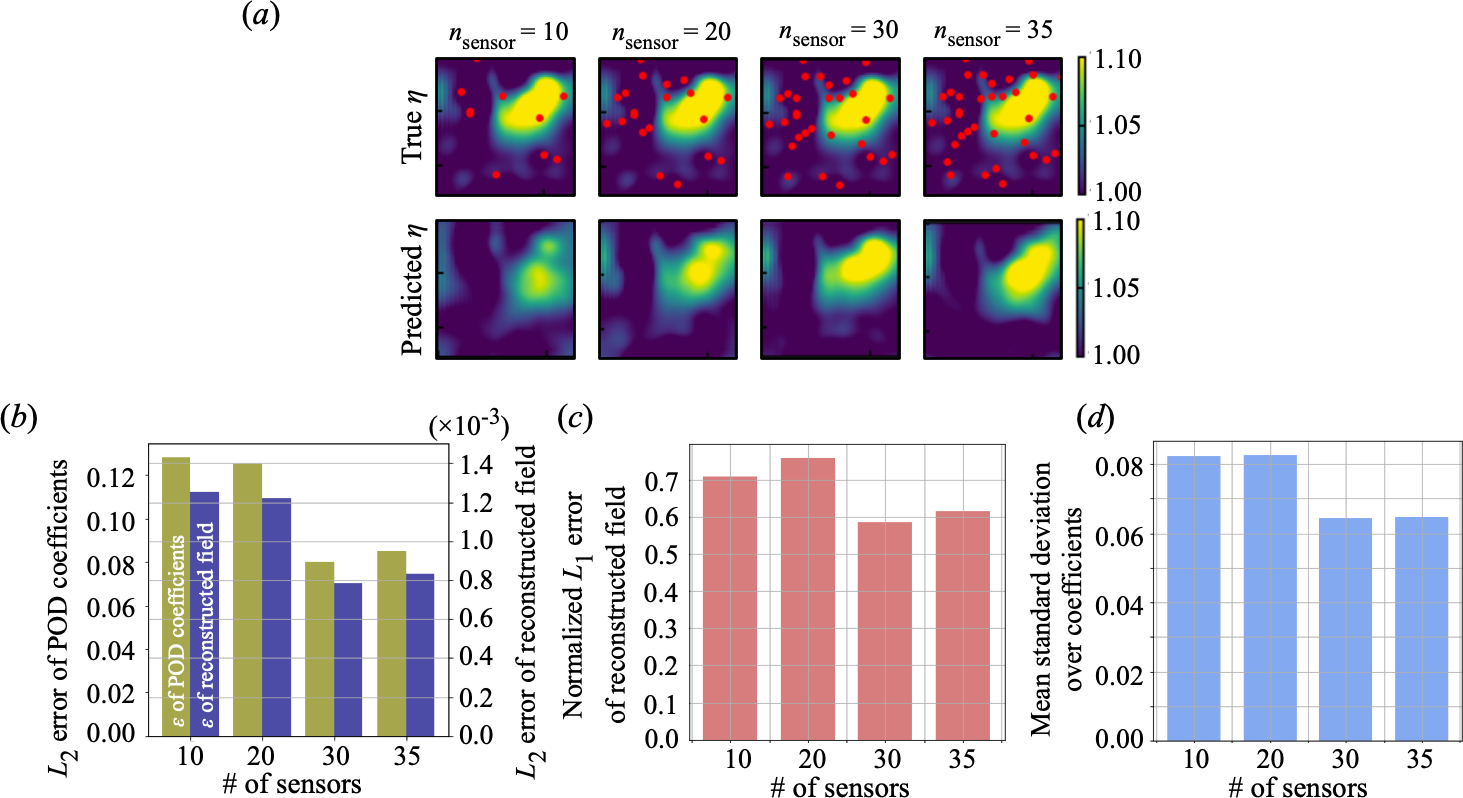}
    \caption{The effect of the number of sensors {for coefficient estimation of shallow water equation} on the accuracy. $(a)$ Field reconstructions with 10, 20, {30}, and 35 sensors. The relationship between the number of sensors and; $(b)$ $L_2$ errors of estimated POD coefficient and reconstructed field; {\color{black}$(c)$ normalized $L_1$ error of reconstructed field;} and {\color{black}$(d)$} ensemble average of estimated standard deviation taking over all POD coefficients. {For reference, the average range of the POD coefficient magnitude was 3.289.}}
    \label{fig:recon}
\end{figure}

Figure \ref{fig:recon} shows the ability of the proposed framework to recover the coherent structures in the field for the shallow water equation system introduced in section \ref{SWE_ROM}. Note that we present $L_2$ error norms of both estimated POD coefficients and reconstructed fields for figure \ref{fig:recon}$(b)$. {For reference, the average range of the 40 POD coefficients magnitude was 3.289 - and that the root mean squared error of the POD coefficients is well-within this range.}
A representative test simulation shows that increasing the number of sensors leads to reduction in errors. Note that due to the random nature of sensor placement, at lower numbers, if point signals of coherent structures are not sampled effectively larger errors may be obtained. This may explain why the lowest $L_2$ error norms appear at 30 sensors as presented in figure{\color{black}s} \ref{fig:recon}$(b)$ {\color{black}and $(c)$}. 
This suggests that an optimal selection of sensor locations {based on regions of high uncertainty} may improve convergence significantly. This inference can be also applied to the estimated standard deviations which do not show significant difference over the considered noise magnitude, as shown in figure \ref{fig:recon}$(d)$.

\begin{figure}
    \centering
    \mbox{
    \subfigure[Test simulation 1]{\includegraphics[width=\textwidth]{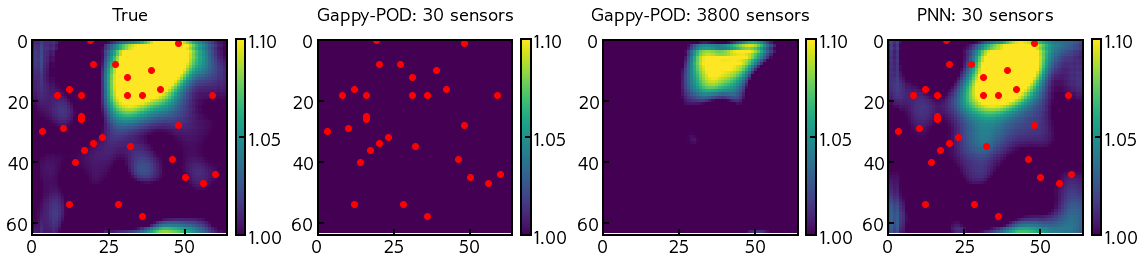}}
    } \\
    \mbox{
    \subfigure[Test simulation 2]{\includegraphics[width=\textwidth]{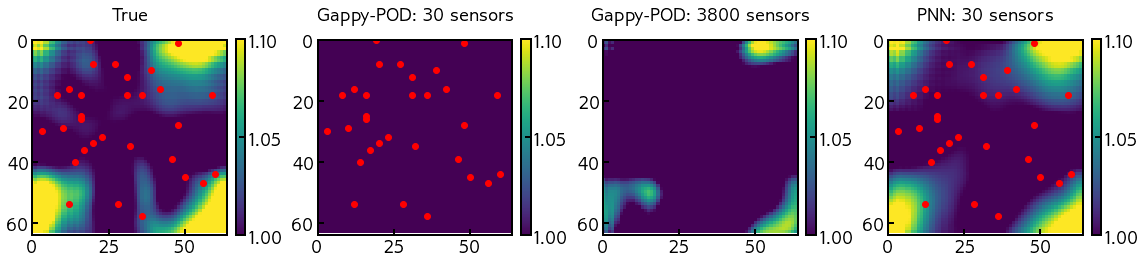}}
    } \\
    \caption{{A comparison between PNN and Gappy POD for POD coefficient reconstruction of two representative testing solutions. For all our testing simulations, {\color{black}the normalized $L_1$ error norms in field reconstruction are 0.155 for the PNN using 30 sensors, 3.18 for Gappy POD using 30 sensors and 0.277 for Gappy POD using 3800 sensors.} Sensor locations (shown in red) are not provided for the validation with 3800 sensors for visual clarity.}}
    \label{fig:Gappy}
\end{figure}

{For a thorough comparison of the PNN reconstruction with the well-known flow-field reconstruction methods, we show a comparison for flow reconstruction using 30 sensors for the Gappy POD method \cite{everson1995karhunen} in figure \ref{fig:Gappy}. The number of sensors (30) is two orders of magnitude lower than the total number of degrees of freedom (4096) and therefore linear reconstruction methods are destined to be at a disadvantage for this test case. We also validate our Gappy POD method by also comparing against improved linear reconstructions utilizing 3800 sensors. 
{\color{black}For all our testing simulations, the normalized $L_1$ error norms in field reconstruction are 0.155 for the PNN using 30 sensors, 3.18 for Gappy POD using 30 sensors and 0.277 for Gappy POD using 3800 sensors. This result represents the advantages of using a nonlinear reconstruction procedure.}
Let us also remind that the Gappy POD reconstruction for POD coefficients is deterministic whereas the PNN provides uncertainty predictions conditioned on the training data. The latter aspect will be explored in greater detail when the PNN is reformulated for direct spatial field recovery for challenging engineering and geophysical applications.}

\begin{figure}
    \centering
    \includegraphics[width=0.9\textwidth]{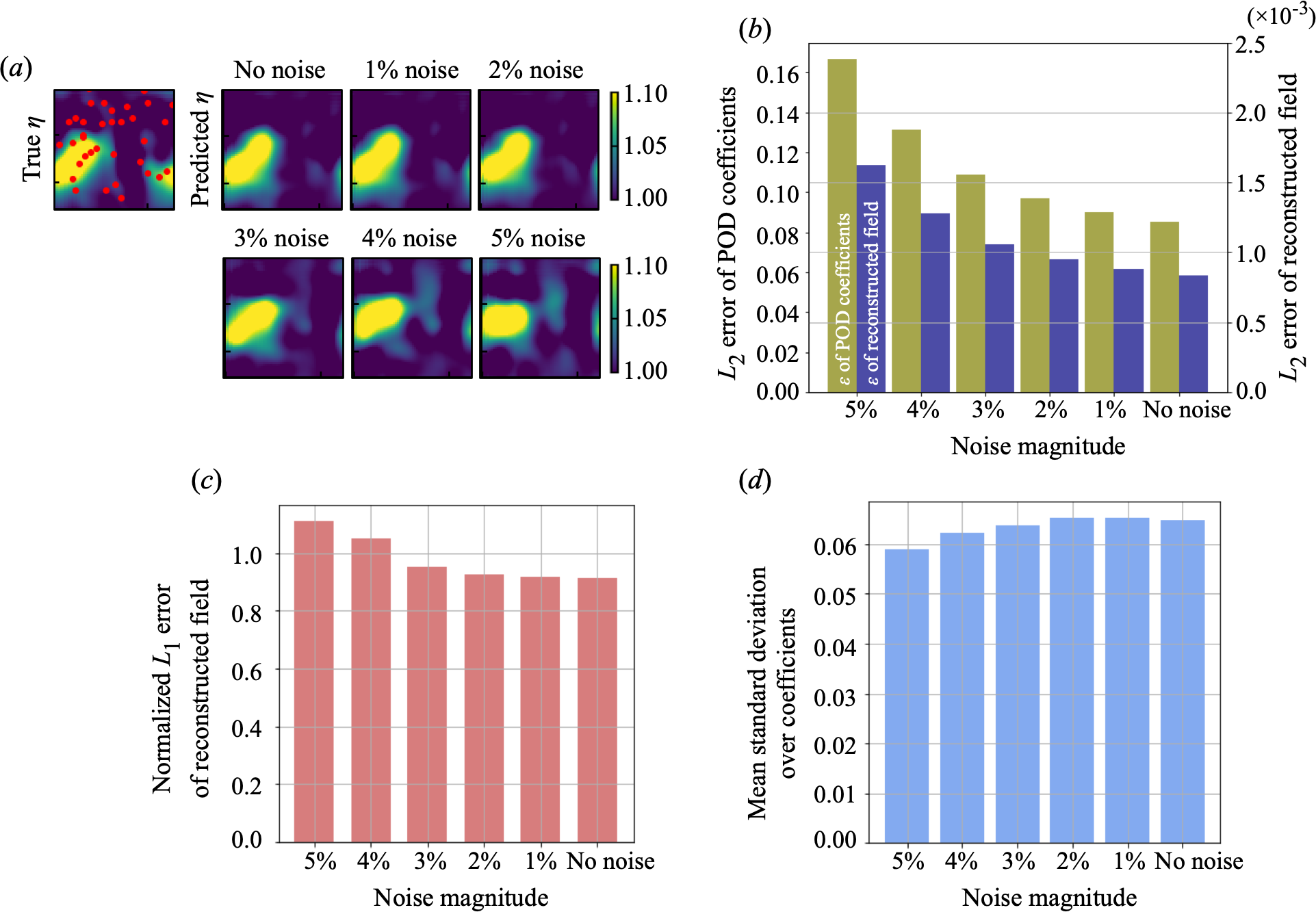}
    \caption{Dependence of accuracy {for coefficient estimation of shallow water equation} on the magnitude of noisy input. $(a)$ Field reconstructions with 35 random sensors. The relationship between the number of sensors and; $(b)$ $L_2$ errors of estimated POD coefficient and reconstructed field; {\color{black}$(c)$ normalized $L_1$ error of reconstructed field;} and {\color{black}$(d)$} ensemble average of estimated standard deviation taking over all POD coefficients. {For reference, the average range of the POD coefficient magnitude was 3.289.}}
    \label{fig:noise}
\end{figure}

Following our assessments with reconstruction without the presence of sensor noise, we turn our attention to the effect of measurement noise at the random sampled locations looking at 35 randomly placed sensor locations. Results from training on noisy inputs are shown in figure \ref{fig:noise}. Figure{\color{black}s} \ref{fig:noise}$(b)$ {\color{black}and $(c)$} indicate that the framework becomes unsuitable for inputs perturbed with uniformly sampled noise corresponding to around 4\% of the maximum value of the field. This may be due to the relatively low difference in magnitudes of the coherent structures in the field from the background flow. 
With regard to estimated standard deviations shown in figure \ref{fig:noise}{\color{black}$(d)$}, we clarify that the uncertainty estimation for the POD coefficients do not show significant difference over the considered noise magnitude. We should note that this is because the estimated confidence interval is just for estimations and not a barometer for error. The reader must note that perturbation of inputs by noise causes errors but does not significantly affect the posterior which assumes that the input is also a random variable. For both assessments in figures \ref{fig:recon} and \ref{fig:noise}, this also indicates that a greater number of training snapshots is necessary to reduce the estimated uncertainty. The effect of number training snapshots will be assessed for a different problem later on in this study. 
{Note that we have already assessed NOAA-SST coefficient estimations via the PNN given field measurements so a separate assessment is superfluous here.}

\subsection{Spatial fluid data recovery}
\label{sec:datarecover}

In this section, we introduce the application of the PNN to spatial fluid flow reconstructions with connections to real engineering and geophysical applications.
We consider a two-dimensional cylinder wake (sec. \ref{sec:tcw}), the wake of a NACA0012 airfoil with a Gurney flap (sec. \ref{sec:ngf}), and the NOAA Optimum Interpolation Sea Surface Temperature data set (sec. \ref{sec:noaasst}). We would like to note that the latter is constructed from satellite and ship observations and represents a real-world flow field reconstruction task with no underlying information of governing equations. Another important remark here is that while previous flow-field reconstructions were via POD coefficient estimation, the PNN is 

For the covered examples in this section, we consider two types of problems: (1) estimation of sensor measurement $s_{\rm target}$ from other sensors ${\bm s}_{ \rm input}$, and (2) estimation of whole flow field ${\bm z}$ from local sensor measurements ${\bm s}_{\rm input}$.
These settings can be expressed as
\begin{align}
\{\pi({{\bm s}_{\rm input}}),\mu({{\bm s}_{\rm input}}),\sigma({{\bm s}_{\rm input}})\}={\cal F}({{\bm s}_{\rm input}}),~~~p({s}_{\rm target} | {\bm s}_{\rm input}) = \sum_{i=1}^{m}\pi_i ({\bm s}_{\rm input}) \mathcal{N}(\mu_i ({{\bm s}_{\rm input}}),\sigma_i ({\bm s}_{\rm input})),\label{eq:ss}\\
\{\pi({{\bm s}_{\rm input}}),\mu({{\bm s}_{\rm input}}),\sigma({{\bm s}_{\rm input}})\}={\cal F}({{\bm s}_{\rm input}}),~~~p({\bm z}| {\bm s}_{\rm input}) = \sum_{i=1}^{m}\pi_i ({\bm s}_{\rm input}) \mathcal{N}(\mu_i ({{\bm s}_{\rm input}}),\sigma_i ({\bm s}_{\rm input}))\label{eq:sw},
\end{align}
for the sensor estimation and the whole field estimation, respectively.

\subsubsection{Two-dimensional cylinder wake}
\label{sec:tcw}

Let us consider a two-dimensional cylinder wake at ${\rm Re}_D=100$ as an example of application to unsteady flows around a bluff body. The data set has been obtained by using a two-dimensional direct numerical simulation (DNS) \cite{TC2007,CT2008}. The governing equations are the incompressible Navier--Stokes equations,
\begin{eqnarray}
\nabla \cdot {\bm u}= 0, \\
\frac{\partial {\bm u}}{\partial t}+{\bm u} \cdot \nabla {\bm u}=-\nabla p+\frac{1}{{\rm Re}_D}\nabla^2\bm u,
\end{eqnarray}
where $\bm u$, $p$ and ${\rm Re}_D$ are the non-dimensionalized velocity vector, pressure and Reynolds number based on the cylinder diameter $D$, respectively.  
Five nested levels of multi-domains are considered for numerical set up.
The finest level of the considered domain here is ($x/D$, $y/D$) = $[-1, 15] \times [-8, 18]$ and the largest domain is ($x/D$, $y/D$) = $[-5, 75] \times [-40, 40]$.  The time step for present DNS is $\Delta t=2.50\times 10^{-3}$.  
As for the training data set, the domain around a cylinder body is extracted, i.e., ($(x/D)^*$, $(y/D)^*$) = $[-0.7, 15] \times [-3.3, 3.3]$ and ($N_x$, $N_y$) = (192, 70). The vorticity field is used as both input and output attributes in this case. Five sensor measurements located on a cylinder surface are chosen as input data ${\bm s}_{\rm input}$ for both the sensor estimation and the whole field reconstruction. For both problem settings with the cylinder data set, we prepare 100 snapshots over approximately 4 periods in time as the training data.
The assessments are conducted using test data set which also contains 100 snapshots over approximately 4 periods and excludes from the training process.

\begin{figure}[t]
    \centering
    \includegraphics[width=0.75\textwidth]{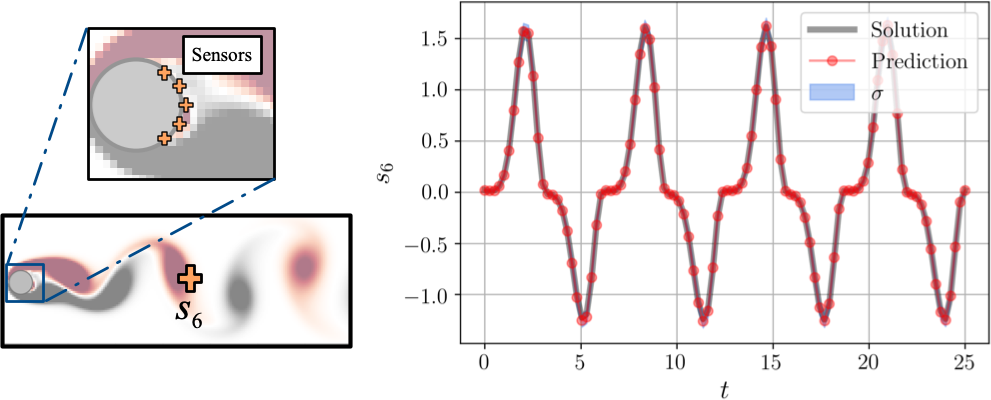}
    \caption{Sensor estimation of cylinder wake using the PNN. The left side shows both input and output sensor locations. $L_2$ error norm $\epsilon=||s_{6{\rm ,DNS}}-s_{6{\rm ,ML}}||_2/||s_{6{\rm ,DNS}}||_2$ is 0.0137. Note here that the $\sigma$ can be hardly seen due to an excellent agreement of estimation with the reference.}
    \label{fig:ss_cy}
\end{figure}

For the problem settings of spatial fluid data recovery, we first apply the PNN to estimate local sensor measurements, and then extend it to the concept of whole flow reconstruction.
Here, let us present the result of estimation on sensor 6 $s_6$ located on a wake region from the sensors located on cylinder surface in figure \ref{fig:ss_cy}.
The estimations (red circles) are in excellent agreement with the reference data.
This trend can be also seen from a quantitative assessment with $L_2$ error norm $\epsilon=||s_{6{\rm ,DNS}}-s_{6{\rm ,ML}}||_2/||s_{6{\rm ,DNS}}||_2$ of 0.0137.
Since the cylinder wake at the present Reynolds number is on periodic nature in time, high standard deviation regions are not observed. This observation also enables us to have confidence for this estimation.

\begin{figure}[t]
    \centering
    \includegraphics[width=\textwidth]{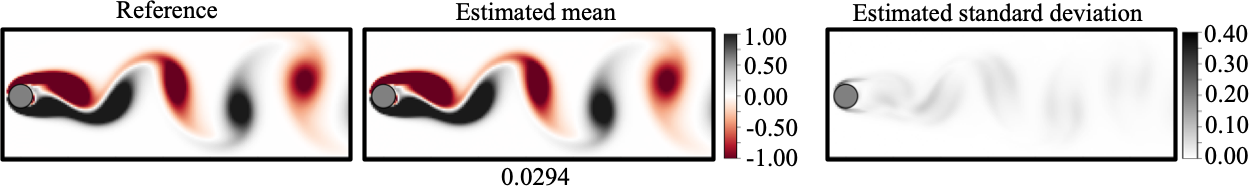}
    \caption{Wake reconstruction of cylinder flow using the PNN. The value below the estimated mean field $\mu$ indicates $L_2$ error norm $\epsilon=||\omega_{{\rm DNS}}-\omega_{{\rm ML}}||_2/||\omega_{{\rm DNS}}||_2$.}
    \label{fig:sw_cy}
\end{figure}

We then extend the model to the reconstruction of the whole wake, as shown in figure \ref{fig:sw_cy}. Analogous to the local sensor estimation, we see that the estimated flow field shows nice agreement with the reference DNS data from both qualitative and quantitative assessments. What is notable is that the PNN provides us with the standard deviations of estimation as shown on the right side of figure \ref{fig:sw_cy}.
In this case, the probabilistic machine learning model tells us that the standard deviation on regions of vortex shedding and separated shear layer is relatively larger than that on other portions. This observation coincides with the fact that these regions have higher fluctuation than the other regions without vortex shedding.

\subsubsection{Wake of NACA0012 airfoil with a Gurney flap}
\label{sec:ngf}

Next, we consider the complex two-dimensional wake behind a NACA0012 airfoil with a Gurney flap. The flow field is also periodic in time analogous to the cylinder problem. However, the wake here is comprised of multiple dominant frequencies.
The data set is generated using two-dimensional DNS at ${\rm Re}_c =1000$, where $c$ is the chord length \cite{GMTA2018}. It is known that various types of wakes can emerge depending on the angle of attack $\alpha$ and the Gurney-flap height $h$ \cite{GMTA2018}. In the present paper, the case of $h/c=0.1$ with $\alpha = 20^\circ$ which exhibits 2P wake is chosen for demonstration, as shown in figure \ref{fig:dataset}$(b)$. For the numerical setup, five nested levels of multi-domains are considered as well as the cylinder problem. The finest domain range is ($x/c$, $y/c$) = $[-1, 1] \times [-1, 1]$ and the largest domain is ($x/c$, $y/c$) = $[-16, 16] \times [-16, 16]$.  The time step is $\Delta t=10^{-3}$. The size of utilized domain and the number of grid points for data set are $[-0.5, 7] \times [-2.5, 2.5]$ and $(N_x^*, N_y^*) = (352, 240)$. We use the vorticity field $\omega$ as the input and output attributes. Five sensor measurements located on surface of an airfoil are chosen as input data for both the sensor estimation and the whole field reconstruction. The number of snapshots $n_{\rm snapshot}$ used for the baseline model are the same as that in the cylinder problem such that 100 snapshots over approximately 4 periods for both training and test data. Note that we also investigate the dependence on the number of the snapshots using the example of NACA0012 wake.

\begin{figure}
    \centering
    \includegraphics[width=0.65\textwidth]{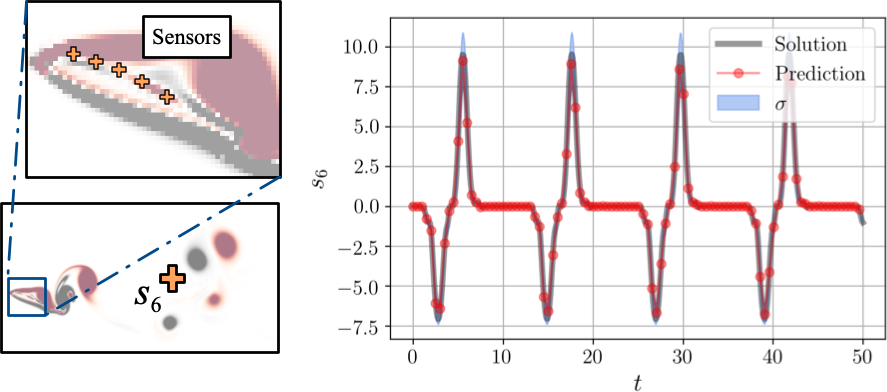}
    \caption{Sensor estimation of NACA0012 wake with a Gurney flap using the PNN. The left side shows both input and output sensor locations. $L_2$ error norm $\epsilon=||s_{6{\rm ,DNS}}-s_{6{\rm ,ML}}||_2/||s_{6{\rm ,DNS}}||_2$ is 0.0429.}
    \label{fig:ss_af}
\end{figure}

The PNN is applied to the local sensor estimation of NACA0012 wake, as shown in figure \ref{fig:ss_af}. As it can be seen, the estimated plots are in great agreement with the reference data. Regions of noticeable standard deviation can be seen near the peaks although these were not observed with sensor estimation for the cylinder problem. This indicates the true peaks in the curve may show some variation from the predicted values given the training of the network. This is due to the difference in the complexity of the flows, i.e., the contained frequency contents on the wakes, as mentioned above. 

\begin{figure}
    \centering
    \includegraphics[width=0.9\textwidth]{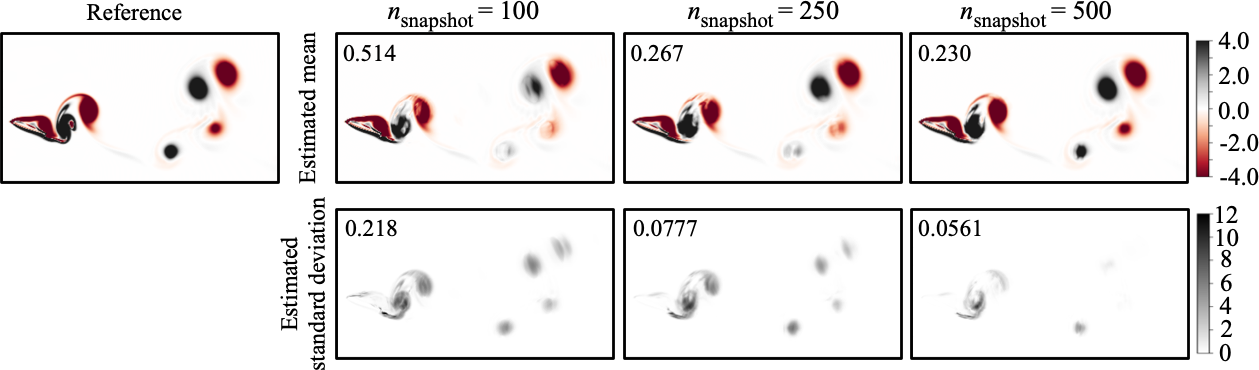}
    \caption{Wake reconstruction of NACA0012 with a Gurney flap depending on the number of the snapshots for training. The values inside the estimated mean field indicates $L_2$ error norm $\epsilon=||\omega_{{\rm DNS}}-\omega_{{\rm  ML}}||_2/||\omega_{{\rm DNS}}||_2$. The values inside the estimated standard deviation express the ensemble-averaged standard deviation over the field.}
    \label{fig:sw_af}
\end{figure}

The estimated whole fields from the sensor measurements located on the surface of airfoil are shown in figure \ref{fig:sw_af}. With $n_{\rm snapshot}=100$, the location and size of vortex structures are captured reasonably well by using the present model. However, the $L_2$ error norm $\epsilon=||\omega_{{\rm DNS}}-\omega_{{\rm ML}}||_2/||\omega_{{\rm DNS}}||_2$ is 0.514, which is substantially larger than that in the case of cylinder wake, despite the reasonable estimation. This is because the $L_2$ error norm is known as a strict measurement of difference and does not account for translational or rotational similarities \cite{FFT2020,FFT2019a}. With regard to the standard deviation distribution, the low confidence interval region are concentrated in the regions of vortical structures, similar to the cylinder problem. The dependence on the number of snapshots used for training is also examined considering $n_{\rm snapshot}=\{100~{({\rm baseline}),~250, ~500}\}$ as shown in figure \ref{fig:sw_af}. We would like to draw attention to the fact that regions with high uncertainty shrink with the increasing the number of snapshots. This highlights the true benefit of the PNN; namely, the better estimate due to the improvement in training data (i.e., greater number of snapshots) is properly quantified as an increased confidence.  These results with the NACA0012 airfoil suggest that the present probabilistic framework can perform well in reconstructing complex flows while providing feedback about the viability of the training data for learning.

\begin{figure}
    \centering
    \includegraphics[width=0.8\textwidth]{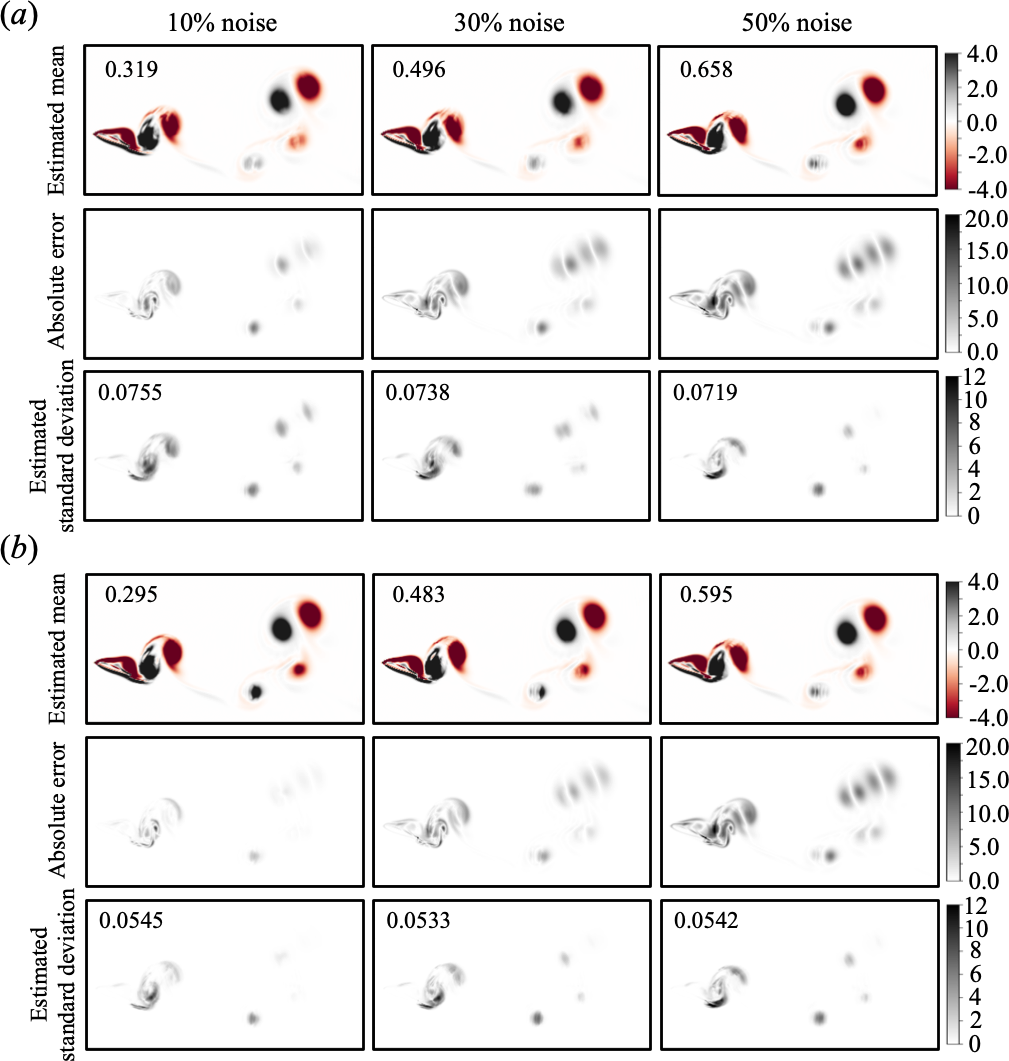}
    \caption{Dependence of estimation accuracy for the whole {wake} field { of NACA0012 with a Gurney flap} on the magnitude of noisy input using the machine-learned model with $(a)$ $n_{\rm snapshot}=250$ and $(b)$ $n_{\rm  snapshot}=500$. The values inside the estimated mean field indicates $L_2$ error norm. The values inside the estimated standard deviation express the ensemble-averaged standard deviation over the field.}
    \label{fig:af_noise}
\end{figure}

Next, let us examine the robustness for noisy input in figure \ref{fig:af_noise}, which is analogous to the assessment with the shallow water equation presented in figure \ref{fig:noise}. 
{The noisy input is designed with Gaussian distribution based on the ratio, i.e., 10, 30, and 50\%, for the maximum absolute value of the field.}
Adding noise to the input leads to increasing the error as can be seen by both $L_2$ error norms and absolute error maps in figure \ref{fig:noise}. Note that the standard deviations have no significant difference among the considered noise level with same number of snapshots due to the fact that the estimated confidence interval is not a barometer for error. This trends coincide with the example of shallow water equation shown in figure \ref{fig:noise}. By comparing figures \ref{fig:af_noise}$(a)$ and $(b)$, we can find that the error decreases by increasing the number of training snapshots, as we expected from the observation with the example of shallow water equation in figure \ref{fig:noise}. We reiterate, therefore, that the proposed formulation does not absolve the user of the best-practices of ML methods (for instance introduced in \cite{FFT2020}). Adequate validation with held-out testing data sets cannot be precluded for the purpose of error-diagnostics. However, the effect of improved learning can be directly correlated with the physics of the predictive task --- in this case flow field recovery. 

\subsubsection{NOAA sea surface temperature}
\label{sec:noaasst}

\begin{figure}[t]
    \centering
    \includegraphics[width=0.5\textwidth]{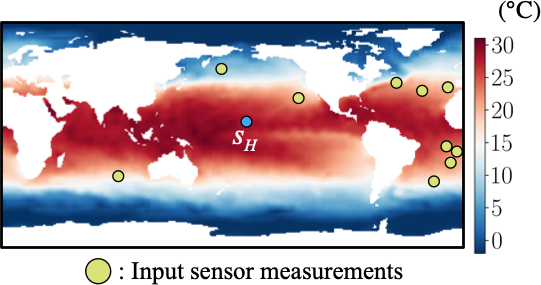}
    \caption{Instantaneous temperature field on the sea surface. Green circles are used sensors as input data.}
    \label{fig:sst_ps}
\end{figure}

\begin{figure}
    \centering
    \includegraphics[width=0.6\textwidth]{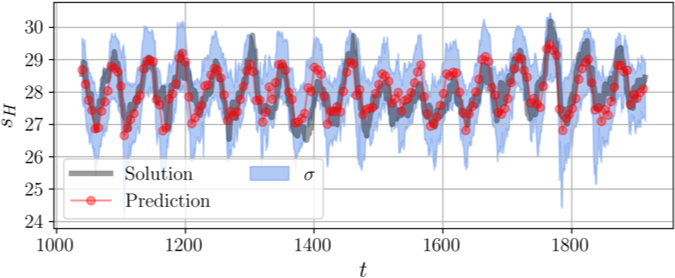}
    \caption{Sensor estimation of sea surface temperature using the PNN. $L_2$ error norm $\epsilon=||s_{H{\rm ,Ref}}-s_{H{\rm ,ML}}||_2/||s_{H{\rm ,Ref}}||_2$ is 0.0163.}
    \label{fig:ss_sst}
\end{figure}

{We here consider the NOAA sea surface temperature data set so as to assess the applicability of the PNN to practical applications of spatial field recovery.
The set-up for training and test data is basically same as that for the parametric surrogate discussed in section \ref{sec:3-1-2}.
Analogous to them, the input sensors for the baseline model are placed from the region of $50{^\circ}$ S to $50{^\circ}$ N as shown in figure \ref{fig:sst_ps}.}
We present the result of local sensor estimation from 10 other sensors in figure \ref{fig:ss_sst}. The reader may observe that the machine learning model can capture the seasonal periodicity quite well. What is notable here is that the interval of standard deviation is wider than that with cases of numerical simulation data, i.e., cylinder wake and NACA0012 airfoil with a Gurney flap. One of possible reasons is the complexity environmental processes which have been added on seasonal temperature variance, e.g., global warming, since the test data is set as extrapolation in time against the used range for training.

\begin{figure}
    \centering
    \includegraphics[width=1.0\textwidth]{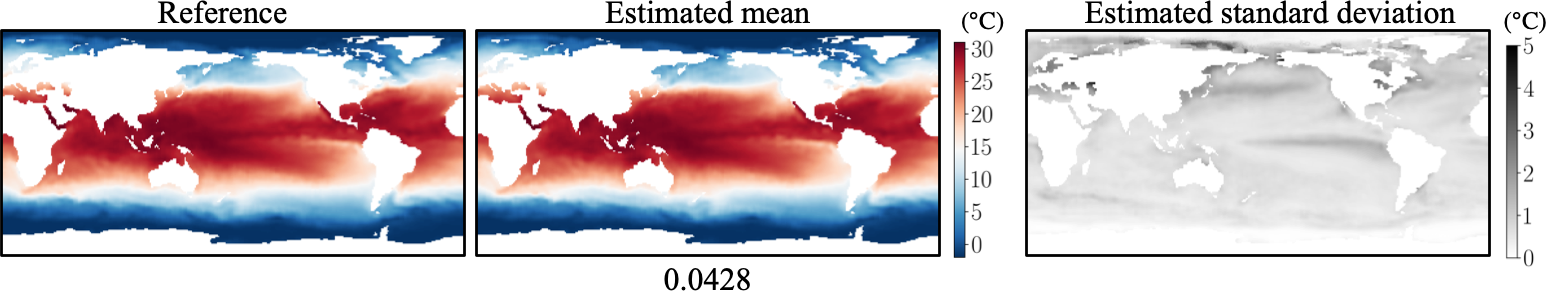}
    \caption{Whole field reconstruction of sea surface temperature. {Note that the estimated standard deviation is the time-ensemble average value.} The value below the estimated mean field indicates $L_2$ error norm $\epsilon=||T_{{\rm Ref}}-T_{{\rm ML}}||_2/||T_{{\rm Ref}}||_2$.}
    \label{fig:sw_sst}
\end{figure}

Let us also consider to the estimation of global temperature field, as shown in figure \ref{fig:sw_sst}. The reconstructed field shows reasonable match with the reference field and this can be also found by $L_2$ norm assessment. The benefit of the proposed framework can be seen when the variance predicted by the framework is correlated to the global grid. We note that the presence of uncertainty is most likely influenced by the choice of input sensor locations. We can also see relatively higher standard deviation region across the Pacific Ocean near South America. This is because of El Ni\~{n}o (sea temperature is higher than usual) and La Ni\~{n}a (lower than usual), which emerge every few years in this area \cite{KFB2016}. Hence, the temperature fluctuation in this area is larger than that in the other areas.

\begin{figure}
    \centering
    \includegraphics[width=1.0\textwidth]{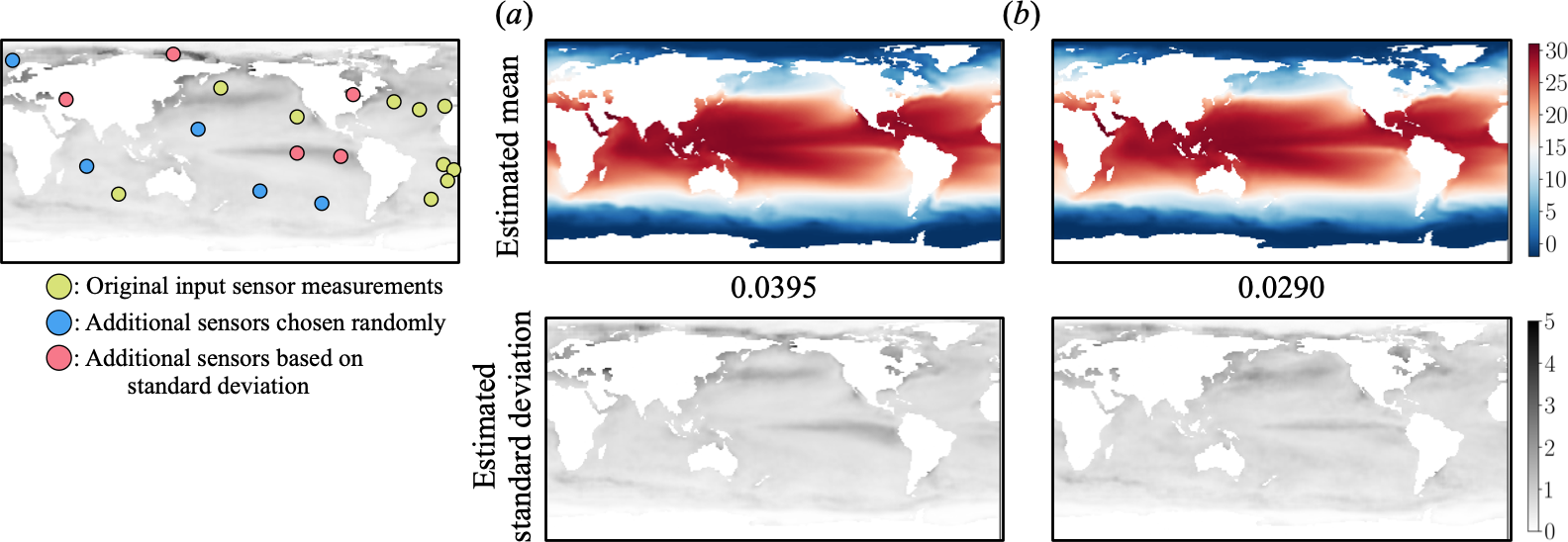}
    \caption{Influence of estimation accuracy {for the whole field of sea surface temperature} on additional sensors. Added 5 sensors $(a)$ randomly and $(b)$ based on the estimated standard deviation in figure \ref{fig:sw_sst}. The values underneath the estimated mean fields indicate $L_2$ error norm.}
    \label{fig:15sens}
\end{figure}

The results of the present study educate us about the possible applications of probabilistic neural networks so that we may be able to assess not only reliability of estimated results but also the characteristics of given training data set by focusing on estimated uncertainties. Furthermore, we can also utilize the estimated field to place additional sensors efficiently since we can observe high standard deviation areas through our probabilistic predictions. Here, let us also assess this viewpoint by putting five additional sensors, as shown in figure \ref{fig:15sens}. We consider two cases by adding 5 sensors $(a)$ randomly and $(b)$ based on the estimated standard deviation in figure \ref{fig:sw_sst}.
As presented, the latter outperforms the case of randomly chosen sensors in terms of $L_2$ error norm. In addition, the sensor selection based on the estimated confidence interval leads to lower standard deviation, e.g., on the Pacific Ocean near South America where El Ni\~{n}o and La Ni\~{n}a can be observed as mentioned above. Through the results in our test cases, we see great potential of the present probabilistic framework for various studies in fluid dynamics.

\section{Concluding remarks}

In this investigation, we have introduced probabilistic neural networks for addressing questions related to the uncertainty quantification of data-driven surrogate model applications to fluid flows. Probabilistic neural networks bestow the ability to parameterize the output as a sample from a Gaussian or a mixture of Gaussian distributions. Consequently, every prediction is accompanied by a confidence interval that may guide the user of the framework about potential errors due to insufficient training data. This represents an improvement on the majority of data-driven studies in fluid mechanics which formulate a deterministic prediction requirement for their surrogates. To demonstrate the viability of the proposed framework, we first deployed it for a parametric model-order reduction task where a surrogate model was constructed for the shallow-water equations. Our surrogate modeling task was particularly complicated by the presence of measurement noise as well as incomplete observations of all relevant conserved variables. The proposed framework was seen to predict, non-intrusively, the evolution of a Gaussian excitation of the field with uncertainty estimates. Following this task, we deployed the framework to a data-recovery problem where measurements at random sensor locations present in a flow-field were used to reconstruct the entire field in the presence of input noise. {We also obtained favorable comparisons for our probabilistic flow-field reconstruction against Gappy POD - a linear reconstruction technique.}

Following assessments for the shallow water equations, the probabilistic neural network was then applied to the spatial reconstruction of flow-fields from data sets with connections to engineering and geophysical applications such as for two-dimensional cylinder flow data, flow around the NACA0012 airfoil with a Gurney flap and for the NOAA optimal interpolation sea-surface temperature data set. {This reconstruction procedure relied on a direct prediction of flow-fields without an intermediate POD-based compression method.} For the problems of two-dimensional cylinder wake and NACA0012 airfoil with a Gurney flap, we found that the present probabilistic model estimated sensor values and reconstructed entire flow fields well while providing uncertainty estimates for the machine-learned estimation. For examples of cases without a modelled governing equation, i.e., the NOAA sea surface temperature data set, our results indicated that the reliability of estimated results as well as the characteristics of a given training data set could be analyzed with the present probabilistic model which shows estimated uncertainties. {We note that the direct reconstruction of spatial fields may lead to the violation of physical laws unless the neural network architectures are actively penalized for these violations via constraints. One approach to deploy this is through the use of PDE constraints as regularization terms in the optimization statement such as the well-known physics-informed neural networks \cite{raissi2017physics}. However, extensions to data sets without knowledge of the underlying PDEs such as NOAA-SST is not trivial and would require other strategies such as data augmentation and constrained network design to satisfy symmetries such as the use of projection-basis networks \cite{ling2015evaluation}. This is an active area of investigation.}

To address the issue of \emph{what} uncertainty is being quantified, we assessed our frameworks for different noise perturbations and training data sizes. We were able to ascertain that the addition of noise to inputs (after training) caused errors in the mean but did not affect confidence intervals. In contrast, improved training data (by utilizing more snapshots) directly led to reducing standard deviations because confidence in outputs was improved. With noisy targets, the framework was adept at characterizing useful confidence intervals. In addition, by plotting standard deviations on the computational grid for the latter case, the regions contributing to the learning difficulty could easily be ascertained for further sampling. Therefore, our recommendation for the use of such frameworks is for applications where both targets and input data are noisy \emph{before} training - a common phenomenon. In the absence of such training data, we suggest the addition of artificial noise (within a certain range) to improve the robustness of the predictions in terms of mean errors. Armed with this knowledge, these probabilistic predictors may be utilized for greater interpretability in data-driven forecasting, reconstruction or model-order reduction tasks with the potential for establishing a feedback loop to improve training data on the fly. We hint towards this using intelligent sampling for the NOAA sea-surface temperature reconstruction task. Notably, the utility of the approximation of the target distribution also has direct applications for the assessment of model quality in the absence of training data.

\textcolor{black}{We remark that the confidence intervals are merely representations of the standard deviation around an expected value, i.e., both the standard deviation and the expectation are learned in a semi-supervised manner to maximize the likelihood of the targets given the inputs. For instance, the greyed out regions of each line plot indicate, with confidence equivalent to one standard deviation, the probability of the true prediction deviating from the mean. Therefore, the viability of using a prediction (in the form of the mean and the intervals) is intimately associated with the target application.} We do not extend the interpretation of the confidence intervals beyond the above characterization, i.e., we do not claim that our probabilistic estimates are the `true' posterior distributions for the targets. Such posterior approximations from an inference method (like variational inference techniques) can only be verified against a fully Bayesian approach, like a Markov Chain Monte Carlo (MCMC) sampling. However for several problems of interest in deep learning, including our applications, doing a full MCMC over all the parameters of the network is prohibitively expensive due the large number of model parameters, i.e., the weights of the network. \textcolor{black}{The departure from being a full posterior arises from the fact 1) The model parameters (such as the weights of the network) are not sampled from a distribution 2) Priors are implicit, i.e., defined by the training set rather than being explicitly defined and 3) The estimated distribution is constrained to be a parametrized conditional density distribution $p({\bm y} | {\bm x})$ as shown in equation \ref{eq:GMM}, instead of a generic posterior distribution. However, this does not mean that the estimated confidence intervals are uninformative. The error bars shown in our results represent the uncertainty quantification in our estimates that arise from the network, under a given training data and a Gaussian approximation. This is clearly seen in our results, where a less-informative training data-priors -- either with a reduced number of snapshots (in figure \ref{fig:sw_af}) in training phase or the number of sensors (in figure \ref{fig:noaa_rom_inpsen}$(a)$) -- results in a broader conditional density estimation $p({\bm y} | {\bm x})$, reflecting a lower confidence in the prediction when training sampling is poor.}

{The PNN is a pragmatic solution for the problem of estimating uncertainty in deep neural networks, although it assumes a Gaussian form for the predictive probability distribution. If the predictions are not expected to be Gaussian distributions (for example, in situations with multiple sub-populations or mixed data), then one may simply use the more generic mixture density networks with an appropriate number of mixture components. We have verified this for the case of shallow water equations, where the addition of components does not make any difference (i.e., all but one mixing coefficients end up with near-zero weight after training) in the reconstruction. This validates the Gaussian assumption and corresponding use of PNNs.}

We have considered the use of probabilistic neural networks with a fully-connected structure. However, the fully-connected model often suffers due to the curse of dimensionality since the number of weights is drastically increased with the connected nodes in the model. This suggests that a convolutional neural network formulation \cite{LBBH1998}, which is able to deal with high-dimensional data through the concept of filter sharing, may be investigated for the next step. In addition, these probabilistic models, which can express confidence intervals for predictions, may also be enhanced if they can be applied efficiently to unstructured data that are seen in various applications for fluid dynamics. To that end we are studying the feasibility of graph neural networks \cite{wu2020comprehensive} and generalized moving least squares \cite{trask2019gmls} frameworks.

\section*{Acknowledgement}
This material is based upon work supported by the U.S. Department of Energy (DOE), Office of Science, Office of Advanced Scientific Computing Research, under Contract~DE-AC02-06CH11357. This research was funded in part and used resources of the Argonne Leadership Computing Facility, which is a DOE Office of Science User Facility supported under Contract DE-AC02-06CH11357. R.M. acknowledges support from the ALCF Margaret Butler Fellowship. K. Fukami and K. Fukagata thank the support from Japan Society for the Promotion of Science (KAKENHI grant number: 18H03758). K.T. acknowledges the generous support from the US Army Research Office (grant number: W911NF-17-1-0118) and US Air Force Office of Scientific Research (grant number: FA9550-16-1-0650). The authors acknowledge M. Gopalakrishnan Meena (University of California, Los Angeles) for sharing his DNS data.

\bibliographystyle{unsrt}  
\bibliography{references}  

\end{document}